\documentclass[manuscript]{acmart}

\usepackage{multirow}
\usepackage{multicol}
\usepackage{arydshln}
\usepackage{longtable}
\usepackage{array}
\usepackage{colortbl}
\usepackage{xcolor}
\usepackage{soul}
\usepackage{changepage}
\usepackage{enumitem}
\usepackage{tabularx}
\usepackage{arydshln}
\usepackage{tikz}

\definecolor{lightblue}{HTML}{b0dcf4} 
\definecolor{lightgreen}{HTML}{d8ecc4} 
\definecolor{lightyellow}{HTML}{f8f4b0} 
\definecolor{lightpurple}{HTML}{decee5} 
\definecolor{DR_yellow}{HTML}{ffde17} 
\definecolor{DR_green}{HTML}{8bc640} 
\definecolor{DR_blue}{HTML}{c2cae4} 
\definecolor{DR_orange}{HTML}{ffe3c8} 
\definecolor{DR_purple}{HTML}{9864ac} 
\definecolor{DR_bluegreen}{HTML}{c9ebe7} 
\definecolor{DR_yellowgreen}{HTML}{f6f6cf} 
\definecolor{panels}{HTML}{9c206e}
\newcommand{\highlight}[2]{\colorbox{#1}{#2}}

\definecolor{mediumgray}{HTML}{808080}

\definecolor{VA}{HTML}{FFDFB5} 
\definecolor{design}{HTML}{F7F8A7}
\sethlcolor{design} 
\newcolumntype{M}{>{\ttfamily\small}p{14cm}}
\newcolumntype{L}{>{\raggedright\arraybackslash}p{2.7cm}}

\definecolor{custompurple}{HTML}{a275ab}
\definecolor{custompink}{HTML}{f27d9b}

\newcolumntype{S}{>{\ttfamily\small}p{8.5cm}}
\newcolumntype{X}{>{\columncolor{VA}\ttfamily\small}p{8.5cm}}

\newcommand{\myc}[2][mediumgray]{%
  \textcolor{#1}{#2}%
}

\def\dream{\emph {DREAM~}}
\def\dreamNospace{\emph {DREAM}}

\newcommand{\lowthr}{2.50}
\newcommand{\highthr}{3.50}
\newcommand{\heat}[1]{%
  \begingroup\def\val{#1}%
  \ifdim \val pt < \lowthr pt
    \cellcolor{red!15}\val%
  \else\ifdim \val pt > \highthr pt
    \cellcolor{green!20}\val%
  \else
    \cellcolor{orange!30}\val%
  \fi\fi\endgroup
}

\newcommand{\susLow}{60}
\newcommand{\susHigh}{70}

\newcommand{\heatSUS}[1]{%
  \begingroup\def\val{#1}%
  \ifdim \val pt < \susLow pt
    \cellcolor{red!15}\val%
  \else\ifdim \val pt < \susHigh pt
    \cellcolor{orange!30}\val%
  \else
    \cellcolor{green!20}\val%
  \fi\fi\endgroup
}

\AtBeginDocument{%
  }

\renewcommand\footnotetextcopyrightpermission[1]{}  
\makeatletter
\AtBeginDocument{%
  \def\ps@firstpagestyle{%
    \def\@oddhead{}\def\@evenhead{}%
    \def\@oddfoot{\hfil\footnotesize\thepage\hfil}%
    \let\@evenfoot\@oddfoot}%
  \def\ps@standardpagestyle{%
    \def\@oddhead{}\def\@evenhead{}%
    \def\@oddfoot{\hfil\footnotesize\thepage\hfil}%
    \let\@evenfoot\@oddfoot}%
}
\makeatother
\begin{document}


\title{Guided Reflection for Personal Sleep Insight in Everyday Sleep Tracking}


\author{Bokyung Kim}
\authornote{Both authors contributed equally to this research.}
\email{bkim94@jhu.edu}
\affiliation{%
  \institution{The Johns Hopkins University}
  \streetaddress{3400 N. Charles St}
  \city{Baltimore}
  \state{Maryland}
  \country{USA}
  \postcode{21218}
}

\author{Amama Mahmood}
\authornotemark[1]
\email{amama.mahmood@jhu.edu}

\affiliation{%
  \institution{The Johns Hopkins University}
  \streetaddress{3400 N. Charles St}
  \city{Baltimore}
  \state{Maryland}
  \country{USA}
  \postcode{21218}
}

\author{Honghao Zhao}
\email{hzhao78@jhu.edu}
\affiliation{%
  \institution{The Johns Hopkins University}
  \streetaddress{3400 N. Charles St}
  \city{Baltimore}
  \state{Maryland}
  \country{USA}
  \postcode{21218}
}

\author{Molly E. Atwood}
\email{matwood4@jhmi.edu}
\affiliation{%
  \institution{Department of Psychiatry and Behavioral Sciences, The Johns Hopkins University School of Medicine}
  \city{Baltimore}
  \state{Maryland}
  \country{USA}
}

\author{Luis F. Buenaver}
\email{lbuenav1@jhmi.edu}
\affiliation{%
  \institution{Department of Psychiatry and Behavioral Sciences, The Johns Hopkins University School of Medicine}
  \city{Baltimore}
  \state{Maryland}
  \country{USA}
}

\author{Ziang Xiao}
\email{ziang.xiao@jhu.edu}
\affiliation{%
  \institution{The Johns Hopkins University}
  \streetaddress{3400 N. Charles St}
  \city{Baltimore}
  \state{Maryland}
  \country{USA}
  \postcode{21218}
}

\author{Chien-Ming Huang}
\email{chienming.huang@jhu.edu}
\affiliation{%
  \institution{The Johns Hopkins University}
  \streetaddress{3400 N. Charles St}
  \city{Baltimore}
  \state{Maryland}
  \country{USA}
  \postcode{21218}
}

\renewcommand{\shortauthors}{Kim et al.}



\begin{abstract}

Digital sleep technologies make tracking accessible, yet users often struggle to interpret what changes in their sleep mean. Behavioral sleep medicine addresses this through guided discovery, helping patients develop personal interpretations rather than simply receiving explanations. To bring this to everyday tracking, we present \dream, an LLM-powered voice assistant that monitors conversational sleep diaries, selectively invites users to interpret meaningful changes, and uses their interpretation to tailor subsequent education. We co-designed \dream with sleep specialists iteratively and evaluated it in a six-week field study ($N=14$) against a generic-education control. \dream participants reported greater personal sleep insight, motivation, and willingness to use the system, and described a clearer rationale for trying strategies. Our findings suggest that guided reflection made participants more active interpreters of their own sleep. Furthermore, we argue that expert involvement is not a single transfer of knowledge but an ongoing process of making tacit judgment explicit and testable.

\end{abstract}




\begin{teaserfigure}
  \includegraphics[width=\textwidth]{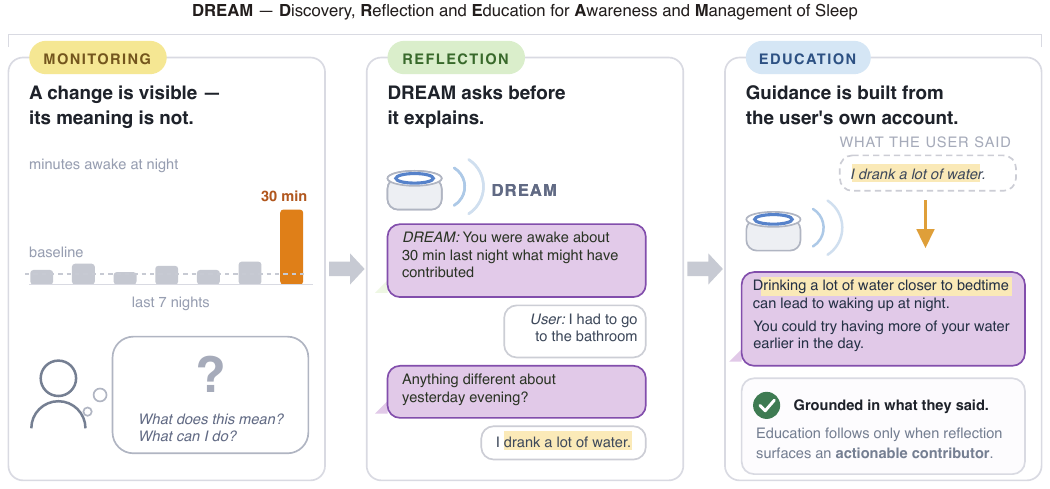}
  \caption{\textbf{D}iscovery, \textbf{R}eflection and \textbf{E}ducation for \textbf{A}wareness and \textbf{M}anagement (\dream) of Sleep connects tracked sleep changes to guidance through the user's
own interpretation. A night deviating from the personal baseline (left) is
brought into reflection, which elicits the user's account rather than
explaining the change (center); education is then built from what that
reflection surfaced and follows only when it identifies a contributor the user
could act on (right). }
    \Description{This figure is titled ``DREAM --- Discovery, Reflection and Education for Awareness and Management of Sleep,'' with the letters D, R, E, A, and M bolded within the words Discovery, Reflection, Education, Awareness, and Management to spell out the acronym. Below the title are three panels connected left to right by two large grey block arrows, each panel headed by a colored pill-shaped label. The first panel, headed ``MONITORING'' in a yellow pill, has the bold heading ``A change is visible --- its meaning is not.'' It shows a bar chart labeled ``minutes awake at night'' with seven bars for the ``last 7 nights''; the first six bars are grey and roughly equal in height, sitting on a dashed ``baseline'' line, while the seventh and final bar is tall and orange, labeled ``30 min'' above it. Below the chart is an outlined icon of a person's head and shoulders, connected by two small dots to a rounded speech box containing a large grey question mark and the italic text ``What does this mean? What can I do?'' The second panel, headed ``REFLECTION'' in a green pill, has the bold heading ``DREAM asks before it explains.'' It shows a small icon of a puck-shaped smart-speaker device with a blue ring around its top rim and curved sound-wave lines beside it, labeled ``DREAM.'' Below is a conversation rendered as chat bubbles: a purple bubble from DREAM reads ``DREAM: You were awake about 30 min last night what might have contributed''; a white outlined bubble reply reads ``User: I had to go to the bathroom''; a second purple DREAM bubble reads ``Anything different about yesterday evening?''; and a second white bubble reply reads ``I drank a lot of water,'' with the phrase ``drank a lot of water'' highlighted in yellow. The third panel, headed ``EDUCATION'' in a blue pill, has the bold heading ``Guidance is built from the user's own account.'' At the top is a grey label ``WHAT THE USER SAID'' above a dashed-outline box repeating the italic text ``I drank a lot of water,'' again with ``a lot of water'' highlighted in yellow. Below that is the same smart-speaker icon with sound-wave lines, and an orange downward arrow leading to a purple box of guidance text: ``Drinking a lot of water closer to bedtime can lead to waking up at night. You could try having more of your water earlier in the day,'' with the opening phrase ``Drinking a lot of water closer to bedtime'' highlighted in yellow. Beneath this box is a green circular checkmark icon next to the bold text ``Grounded in what they said,'' followed by the caption ``Education follows only when reflection surfaces an actionable contributor,'' with ``actionable contributor'' in bold. Together, the three panels illustrate DREAM's pipeline: detecting a meaningful change in sleep data, asking the user reflective questions to surface a likely contributor before explaining anything, and then generating guidance that is explicitly grounded in what the user said, rather than generic advice.}
  \label{fig:teaser}
\end{teaserfigure}


\maketitle
\pagestyle{plain}
\thispagestyle{plain}

\section{Introduction}

Sleep health is an important component of overall health and well-being, with insufficient or disrupted sleep associated with cardiovascular and metabolic health risks, poorer mental health, and reduced daytime performance and productivity~\cite{altevogt2006sleep,buysse2014sleephealth}. Digital sleep technologies have made it easier to monitor sleep and increasingly provide personalized feedback and recommendations~\cite{liang2016sleepexplorer,daskalova2016sleepcoacher,wang2025healthguru}. Yet making changes in sleep visible does not necessarily help people understand what those changes mean in the context of their own lives. People may notice that they slept less than usual or woke repeatedly while remaining unsure what may have contributed or how they should respond~\cite{ravichandran2017making,liang2016sleeptracking,purnell2023wearable}. The challenge is therefore not only noticing a change, but bridging the interpretive gap between that observation and knowing what to do next by understanding it in the context of one's own experience.

Personal informatics research identifies reflection as a process through which people relate personal data to lived experience and develop self-knowledge that can inform action~\cite{li2010stage,epstein2015lived}. In everyday self-tracking, however, much of this interpretation is still left to the user, and tracking does not consistently lead to insight or behavior change~\cite{kerstenvandijk2017personal,cho2022reflection}. Behavioral sleep medicine offers a model for supporting this process more actively. Cognitive Behavioral Therapy for Insomnia (CBT-I), a first-line treatment for chronic insomnia with approximately 70--80\% of patients experiencing treatment benefit~\cite{morin2012chronic,edinger2021guideline}, uses longitudinal sleep diaries to examine patterns and possible contributors and adapt strategies as understanding develops~\cite{walker2022cbti,edinger2021guideline}. Through guided discovery, clinicians use collaborative questioning to help patients examine their own observations and develop interpretations rather than simply receive an explanation~\cite{padesky1993socratic,overholser2011guided}.

This clinical model of collaborative sensemaking demonstrates that guided reflection is critical for translating tracked data into actionable insight. This need extends well beyond diagnosed insomnia. An estimated 33--50\% of adults experience regular sleep difficulties~\cite{rossman2019cognitive} and actively seek to understand and improve their sleep~\cite{murawski2018systematic}, while access to behavioral sleep medicine remains limited, with most large U.S. cities lacking a trained provider~\cite{thomas2016providers}. This raises a question of how such an interpretive approach might support people who monitor and seek to improve their sleep in everyday settings, without attempting to reproduce clinical care. We therefore draw on the principles of guided discovery and sleep-diary review strictly as design resources to construct an interpretive layer that helps general users connect tracking data to their lived experiences.

In digital settings, sleep systems have sought to move beyond monitoring by helping people understand and improve their sleep. Some support reflection through visual exploration and structured prompts~\cite{liang2016sleepexplorer,ploderer2023sleepbeta,liu2020slumberbot}, while others personalize guidance using tracked data and context~\cite{daskalova2016sleepcoacher,wang2025healthguru}. However, these processes are often treated separately, with guidance generated primarily from tracked data and context rather than from the interpretation a user develops about a change in their own sleep. Large language models (LLMs) create opportunities to connect these processes conversationally by bringing personal data into discussion, eliciting users' experiences, and adapting subsequent questions and guidance~\cite{stromel2024narrating,loerakker2025giveandtake,bhattacharjee2026generative,jorke2026bloom}. Realizing this potential requires deciding which changes warrant discussion, how to support interpretation, and when that interpretation provides a basis for guidance.

We therefore ask: \textit{How can we design an LLM-powered conversational system, inspired by guided discovery in behavioral sleep medicine, that supports users in interpreting changes in their sleep and connects those interpretations to subsequent guidance?} We developed \textit{\dream} (Discovery, Reflection and Education for Awareness and Management of Sleep, Fig. \ref{fig:teaser}), an LLM-powered, voice-based system that integrates conversational sleep diaries, longitudinal monitoring, guided reflection, and sleep education. \dream selectively brings sleep changes into conversation, asks users what may have contributed, and uses their interpretations to determine whether and what education should follow.

We developed \dream through a five-stage process with behavioral sleep medicine specialists, including a 10-day technology probe with 11 participants and a six-week field study with 14 participants. In the field study, we compared \dream with a Generic Education control approximating non-personalized guidance in consumer sleep applications~\cite{arroyo2022sleepapps,choi2018sleepapps}. \dream participants reported greater personal sleep insight, motivation, and willingness to use the system, showed greater insight depth in open-ended responses, and more often reported trying a sleep-related strategy. Interviews further illustrated how guided reflection helped participants connect sleep changes to their own circumstances and use those interpretations to make subsequent guidance personally meaningful. These findings show how LLM-guided reflection can support users in doing the interpretive work of making personal data meaningful, while using that understanding to shape what support follows.

This work makes three contributions:
\begin{itemize}[leftmargin=*]

\item \textbf{System: \dream.} We present an LLM-powered, voice-based system that connects longitudinal sleep monitoring, guided reflection, and education grounded in users' interpretations of their sleep.

\item \textbf{Empirical: Initial Evidence on Guided Reflection in Everyday Sleep Tracking.} Through a formative technology probe and a six-week field study, we show how guided reflection can support personal sleep understanding, motivation, and strategy uptake by connecting users' interpretations to subsequent guidance.

\item \textbf{Design: Implications for LLM-Powered Behavioral Health Systems.} Drawing on our iterative design process and deployments, we present design implications for translating specialist-guided reflection into everyday systems. We highlight how LLMs can be structured to elicit users' interpretations of their personal data and seamlessly connect those insights to tailored subsequent support.

\end{itemize}




\section{Related Work}


Our work builds on research in sleep monitoring and its interpretive challenges, reflection on personal health data, and personalized and adaptive support in digital sleep systems.

\subsection{Sleep Monitoring and the Interpretive Gap}
Prospective sleep diaries have long been used to monitor sleep across nights, capturing subjective sleep experiences alongside measures such as sleep timing, duration, awakenings, and sleep efficiency~\cite{carney2012consensus}. In everyday settings, consumer wearables, smartphone applications, and electronic diaries have made longitudinal sleep monitoring easier and more accessible~\cite{deZambotti2019wearable,arroyo2022sleepapps}. Some systems also capture contextual factors to help users examine possible relationships between sleep and everyday behaviors or experiences~\cite{choe2015sleeptight,liang2016sleepexplorer,daskalova2016sleepcoacher}. However, access to more detailed sleep records does not inherently clarify their meaning. Studies of consumer sleep tracking have found that users can struggle to interpret sleep metrics, judge their significance, and identify changes they could make~\cite{ravichandran2017making,liang2016sleeptracking}. Wearable users have similarly reported wanting additional help understanding their sleep data and connecting it to practical behavioral changes~\cite{purnell2023wearable}. Sleep specialists have also described benefits for awareness and engagement alongside concerns that consumer technologies may contribute to misconceptions about sleep~\cite{addison2023consumer}. These studies suggest that longitudinal monitoring can make changes in sleep visible, but additional support may be needed to understand what may have contributed and what the user could do in response.

\subsection{Supporting Reflection on Personal Health Data}
Reflection has been supported through approaches such as data exploration and guided questioning in personal informatics and behavioral sleep medicine, and more recently through conversational and LLM-based systems.
\subsubsection{Reflection in Personal Informatics and Behavioral Sleep Medicine}
Personal informatics research identifies reflection as the process through which people relate personal data to lived experience and develop self-knowledge that can inform action~\cite{li2010stage,epstein2015lived}. Prior work identifies temporal perspective, comparison, discovery, and conversation as resources for structuring reflection~\cite{bentvelzen2022revisiting}. Yet tracking does not consistently lead to insight or behavior change, and many applications leave the interpretive work to the user~\cite{kerstenvandijk2017personal,cho2022reflection}. While prior work outlines the conceptual building blocks of reflection, how a system can actively scaffold a user from a specific observation toward a personal interpretation remains underexplored.

Behavioral sleep medicine offers a related model for actively supporting this interpretive process. Cognitive Behavioral Therapy for Insomnia (CBT-I), a first-line treatment for chronic insomnia~\cite{edinger2021guideline,qaseem2016management}, uses sleep-diary review alongside behavioral and cognitive techniques to examine sleep patterns, possible contributors, and strategies over time~\cite{walker2022cbti,edinger2021guideline}, considering recorded patterns in relation to patients' daily experiences, behaviors, and beliefs rather than as self-explanatory. Guided discovery, a cognitive therapy technique used in CBT-I to examine and reevaluate sleep-related thoughts~\cite{belanger2006clinical,morin2011cognitive,walker2022cbti}, supports this through collaborative questioning: clinicians build on what the person has already shared and clarify uncertainty before drawing conclusions, rather than providing a predetermined explanation~\cite{padesky1993socratic,overholser2011guided}. We adapt this questioning stance in \dream to help users interpret changes in a sleep record in relation to their own experiences, rather than to restructure sleep-related beliefs.

\subsubsection{Conversational and LLM-Based Support for Reflection}
Conversational agents have been used to support reflection on personal data and health behavior. Reflection Companion used short adaptive dialogues to help users notice observations in physical-activity data, consider explanations, and identify actions~\cite{kocielnik2018reflection}, while early rule-based CBT chatbots delivered structured exercises through scripted exchanges~\cite{fitzpatrick2017woebot}.

LLMs have expanded what such conversations can do, following a user's account, asking context-sensitive follow-up questions, and adapting across turns. Recent work has used LLMs to generate narratives from tracker data~\cite{stromel2024narrating}, scaffold journaling and self-reflection with adaptive questions~\cite{kim2024mindfuldiary,nepal2024mindscape,song2025exploreself}, support cognitive reframing of negative thoughts~\cite{sharma2023reframing}, and deliver health coaching that draws on user-provided context~\cite{loerakker2025giveandtake,jorke2026bloom,bhattacharjee2026generative}. While these systems can personalize reflection or guidance from user data and context, less work has examined how conversation can help users develop their own understanding of a change in their personal data before guidance is provided.

\subsection{Personalized and Adaptive Support in Digital Sleep Systems}

In digital sleep systems, reflection and guidance have been operationalized through personalized and adaptive approaches that shape both what support is provided and when it is offered.

\subsubsection{Connecting Reflection and Personalized Guidance}
Digital sleep systems have supported reflection through visual exploration, structured prompts, conversation, and behavioral experiments. SleepExplorer lets users examine relationships between sleep and contextual factors~\cite{liang2016sleepexplorer}, SleepBeta combines reflection prompts, visualizations, and behavioral experiments~\cite{ploderer2023sleepbeta}, and SlumberBot asks predefined questions about factors that may have affected sleep quality~\cite{liu2020slumberbot}. These systems create opportunities for users to reflect, but the user's interpretation of a particular observation does not generally determine what support follows. Other systems use tracked and contextual data to personalize guidance. SleepCoacher identifies relationships between sleep behaviors and outcomes and supports self-experiments~\cite{daskalova2016sleepcoacher}, HealthGuru combines wearable and contextual data in an LLM-based multi-agent architecture to provide adaptive recommendations~\cite{wang2025healthguru}, and consumer applications commonly provide sleep knowledge, behavioral instructions, and feedback~\cite{arroyo2022sleepapps,lancaster2023sleepapps}.

Across this work, reflection and personalized guidance are often disconnected: reflection helps users make sense of their data, while guidance is primarily generated from tracked data and context rather than from the interpretation the user develops. Personal informatics research positions reflection as a way to develop self-knowledge that can inform action~\cite{li2010stage}, while tailored health communication research suggests that personally relevant guidance can increase attention and intentions to act~\cite{rimer2006advancing,bol2020tailored}. Connecting these processes can therefore provide a clearer rationale for action by linking what a user understands about a change to guidance relevant to that understanding. Our work makes this connection explicit by using the user's interpretation of a selected sleep change to determine whether guidance is warranted and what it should address.

\subsubsection{Deciding When to Support Reflection}
Supporting reflection during repeated tracking also requires deciding when additional engagement should occur, since frequent or poorly timed prompts can become repetitive or burdensome in daily tracking contexts where routine interactions accumulate~\cite{kocielnik2018reflection}. Behavioral sleep medicine offers a model in which sleep-diary records help identify patterns worth discussing rather than treating every recorded night as requiring reflection~\cite{walker2022cbti,edinger2021guideline}, and Just-In-Time Adaptive Intervention (JITAI) frameworks similarly decide whether and when to provide support based on changing information about the individual, including the option of providing none~\cite{nahumshani2018jitai}. We draw on these perspectives to make reflection selective: longitudinal sleep data identify when an observation may warrant attention, while the subsequent conversation determines how much additional support is needed.

\section*{Research Overview}

Our work investigates how guided reflection and sleep education can be integrated into everyday sleep tracking. We developed \dream, an LLM-powered, voice-based system that uses longitudinal sleep data to support users in interpreting changes in their sleep and provides education when relevant. Our goal was not to reproduce clinician-led CBT-I, provide diagnosis or treatment, or automate clinical interpretation. Instead, we drew on guided discovery and sleep-diary review in behavioral sleep medicine as design resources for helping users examine changes in their sleep, develop their own interpretations, and determine when further guidance may be useful~\cite{padesky1993socratic,overholser2011guided,walker2022cbti,edinger2021guideline}.

\begin{figure*}[t]
\centering
\includegraphics[width=\textwidth]{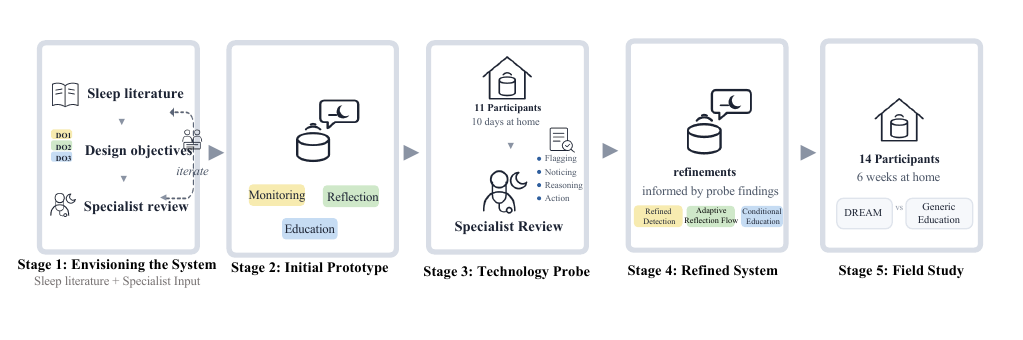}
\caption{Overview of the five-stage design process. Stage~1 translated behavioral sleep medicine practices into an initial design specification with sleep specialists. Stage~2 implemented this specification in an initial prototype. Stage~3 deployed the prototype as a 10-day technology probe and used participant experiences, researcher checks, and specialist review to identify needed refinements. Stage~4 incorporated these refinements into the refined system used in Stage~5, a six-week field study with 14 participants.}
\Description{This figure shows five panels arranged left to right, connected by solid grey triangular arrowheads. Each panel is a rounded rectangle with a thick light-grey border, and each has a bold stage label beneath it. The first panel, labeled ``Stage 1: Envisioning the System'' with the grey subtitle ``Sleep literature + Specialist Input'' beneath it, contains a vertical sequence of three items linked by small grey downward triangles. The first item is an open-book icon next to the bold text ``Sleep literature.'' The second is a stack of three small tags reading ``DO1,'' ``DO2,'' and ``DO3,'' shaded yellow, green, and blue respectively, next to the bold text ``Design objectives.'' The third is an icon of a person wearing a stethoscope with a crescent moon beside them, next to the bold text ``Specialist review.'' Along the right side of this panel, a dashed grey bracket connects ``Specialist review'' back to ``Sleep literature,'' with arrowheads at both ends. The bracket is accompanied by an icon of two people behind a laptop, representing the two sources being compared and discussed, and by the italic grey word ``iterate.'' The second panel, labeled ``Stage 2: Initial Prototype,'' shows a cylindrical smart-speaker icon with a small dome on top, beside a speech bubble containing a crescent moon. Below it, three colored tags sit in a row: ``Monitoring'' in yellow, ``Reflection'' in green, and ``Education'' in blue. The third panel, labeled ``Stage 3: Technology Probe,'' shows at the top an outline of a house with a small cylindrical device inside. Beneath it are the bold text ``11 Participants'' and the regular text ``10 days at home.'' A small grey downward triangle leads to the lower half, where a large icon of a person wearing a stethoscope with a crescent moon sits above the bold text ``Specialist Review.'' To the right of this icon is a document icon with a checkmark badge, representing checks on the collected data. Below it is a bulleted list with blue bullet points reading ``Flagging,'' ``Noticing,'' ``Reasoning,'' and ``Action.'' The fourth panel, labeled ``Stage 4: Refined System,'' repeats the smart-speaker and crescent-moon speech-bubble icon from Stage 2. Beneath it are the bold lowercase text ``refinements'' and the regular text ``informed by probe findings.'' Below these are three colored tags in the same yellow, green, and blue order as in Stage 2: ``Refined Detection,'' ``Adaptive Reflection Flow,'' and ``Conditional Education.'' The fifth panel, labeled ``Stage 5: Field Study,'' shows the same house-with-device icon as Stage 3, above the bold text ``14 Participants'' and the regular text ``6 weeks at home.'' Below are two white rounded boxes with light-grey outlines reading ``DREAM'' and ``Generic education,'' with a small grey ``vs'' between them. Together, the five panels illustrate how the system progressed through five steps. It began with design objectives developed by iteratively comparing and discussing sleep literature and specialist review. These led to a three-component prototype. The prototype was then deployed as an in-home technology probe, whose data was checked and reviewed by specialists. The findings produced a refined system whose three components map by color onto the original ones. Finally, a six-week field study compared the system against generic education.}
\label{fig:design-process}
\end{figure*}

Translating these practices into an everyday system required making explicit decisions that a clinician would otherwise make during an unfolding interaction: which changes in a person's sleep should be brought into discussion, how reflection should proceed, and when guidance should follow. We developed the system iteratively with two behavioral sleep medicine specialists, with 6 and 21 years of clinical experience, through five stages: specialist-informed design, initial implementation, a formative technology probe, system refinement, and a six-week field study (Fig.~\ref{fig:design-process}).

\section{Stage 1: Envisioning the System with Sleep Specialists}

We began by formulating three design objectives based on how behavioral sleep medicine supports the interpretation of longitudinal sleep records. Clinicians selectively review patterns across nights, use guided discovery to help patients develop their own understanding, and connect that understanding to appropriate behavioral guidance~\cite{walker2022cbti,edinger2021guideline,padesky1993socratic,overholser2011guided}. These objectives defined what we wanted the system to support; we then worked with two behavioral sleep medicine specialists to determine how these practices could be translated into an everyday interaction.

\subsection{Design Objectives}
\label{sec:dos}

\begin{enumerate}[leftmargin=*]

\item \textbf{\highlight{lightyellow}{DO1}: Selectively Reviewing Longitudinal Sleep Patterns.}
Clinicians review sleep diaries across sessions and focus discussion on changes and barriers most relevant to the patient's current difficulties~\cite{manber2012dissemination,walker2022cbti}. We therefore sought to initiate reflection selectively from the longitudinal record rather than treat every nightly change as requiring further discussion.

\item \textbf{\highlight{lightgreen}{DO2}: Guiding Discovery of Personal Interpretation.}
Clinicians use guided discovery and Socratic questioning to help patients examine their observations and develop their own understanding~\cite{belanger2006clinical,morin2011cognitive,padesky1993socratic}. We adapted this principle to reflection on sleep data, helping users interpret a change in relation to their own experiences rather than providing an explanation for them.

\item \textbf{\highlight{lightblue}{DO3}: Grounding Behavioral Guidance in Personal Understanding.}
Clinicians use sleep records together with the patient's account and context to determine which behavioral strategies to emphasize~\cite{manber2012dissemination,edinger2021guideline,walker2022cbti}. We therefore sought to ground subsequent guidance in the understanding developed through reflection rather than provide guidance independently of the user's interpretation.

\end{enumerate}

These objectives defined a three-phase workflow corresponding to the design objectives: \textit{monitoring} (\textbf{DO1}) to identify changes worth examining, \textit{reflection} (\textbf{DO2}) to support the user's interpretation of those changes, and \textit{education} (\textbf{DO3}) when that interpretation provided a relevant basis for guidance.

\subsection{Procedure}
We first worked with the specialists to develop the three parts of the initial workflow: (1)~\textbf{Monitoring}: we drafted initial criteria for identifying potentially noteworthy changes in sleep-diary measures based on prior literature~\cite{lichstein2003quantitative,schutterodin2008clinical,akerstedt2018criteria}, then reviewed and refined these criteria with the specialists, along with what sleep and contextual information the morning and evening diaries should capture; (2)~\textbf{Reflection}: we discussed how specialists proceed after identifying a change worth examining and translated the emerging conversational flow into a decision tree; and (3)~\textbf{Education}: we discussed when education should follow reflection, how it should connect to the user's interpretation, and what sleep-hygiene material should ground the educational content.

We then applied the emerging monitoring criteria and conversational flow to simulated sleep cases spanning different sleep changes, user explanations, and levels of uncertainty. With the specialists, we reviewed whether the selected change was appropriate to bring into reflection, how the conversation should proceed, and whether and what guidance should follow. These walkthroughs further refined the initial system logic before implementation.

\subsection{Findings}

The specialist-informed design process resulted in the following decisions for the three phases of the workflow.

\subsubsection{\highlight{lightyellow}{Phase 1: Monitoring}}

\begin{itemize}[leftmargin=*]

\item \textbf{Record sleep outcomes together with daily context.}
In addition to standard sleep-diary measures, the specialists emphasized the importance of capturing contextual information that could help users later interpret a change. We therefore paired the morning sleep diary with an evening diary capturing factors such as stress, mood, caffeine, exercise, and naps.

\item \textbf{Construct a personalized baseline from recent sleep history.}
Using standard sleep-diary measures as a starting point, we worked with the specialists to identify TST, SOL, WASO, SE, NWAK, TIB, and sleep quality as measures for characterizing a person's recent sleep pattern. Rather than interpreting an observation in isolation, these measures were considered relative to the individual's longitudinal record.

\item \textbf{Consider both personal deviation and the magnitude of the sleep issue.}
Specialist discussions highlighted that a change from a person's usual sleep and the magnitude of a sleep problem were not necessarily equivalent. Starting from literature-informed criteria, we refined the measure-specific thresholds with the specialists to distinguish moderate personal deviations, larger deviations, and absolute threshold crossings.

\item \textbf{Prioritize one change for reflection.}
Simulated cases often contained changes across multiple sleep measures. The specialists favored focusing reflection on a single issue rather than raising several changes at once, resulting in the priority order SOL/WASO, followed by TST, TIB, SE, NWAK, and sleep quality.

\end{itemize}

\subsubsection{\highlight{lightgreen}{Phase 2: Reflection}}

\begin{itemize}[leftmargin=*]

\item \textbf{Begin with the user's own interpretation.}
The specialists emphasized first eliciting the user's account rather than presenting a possible explanation for the observed change. Reflection therefore began by asking what the user thought may have contributed.

\item \textbf{Provide progressively more support when the user is uncertain.}
The specialists described becoming more directed only when the user was unable to identify an explanation. We therefore structured reflection to progress from prompting recall of differences in the previous day or routine, to introducing relevant contextual cues, and finally to asking whether the change appeared isolated or recurring.

\end{itemize}

\subsubsection{\highlight{lightblue}{Phase 3: Education}}

\begin{itemize}[leftmargin=*]

\item \textbf{Use reflection to determine what education should address.}
The specialists emphasized that education should not be selected solely from the detected sleep measure. Instead, when reflection surfaced a relevant factor, education connected that factor to the selected sleep issue and offered one or two concrete actions the user could consider.

\item \textbf{Keep guidance grounded in specialist-provided sleep knowledge.}
To constrain the content of these responses, the specialists provided sleep-hygiene material spanning dietary, environmental, behavioral, and emotional factors. This material served as the source for educational responses, while the user's interpretation determined which content was relevant to the current interaction.

\end{itemize}

\section{Stage 2: Initial Prototype}
\label{sec:stage2}

\begin{figure*}[t]
\centering
\includegraphics[width=\textwidth]{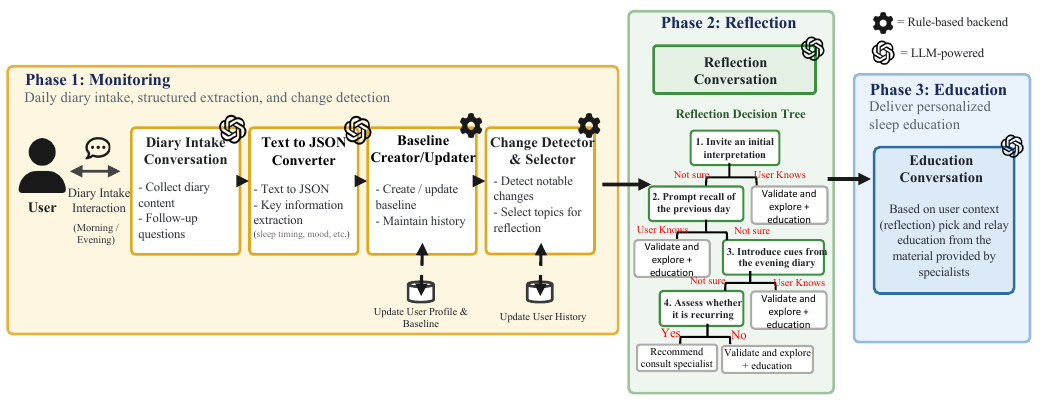}
\caption{System pipeline of the initial prototype. Conversational morning and evening diaries were converted into structured longitudinal records; backend monitoring constructed personal baselines, detected and selected sleep changes, and passed a selected change and relevant context to LLM-guided reflection and education. }
\Description{This figure shows a left-to-right pipeline divided into three phases. Each phase is a rounded rectangle with a colored fill and border, and the phases are connected by thick black arrows. The phases use the same color scheme as the other figures in the paper: yellow for Monitoring, green for Reflection, and blue for Education. The boxes inside each phase have borders in that phase's color, marking which phase each component belongs to. A legend in the upper right defines two icons: a black gear means ``= Rule-based backend,'' and the OpenAI logo means ``= LLM-powered.'' The first and widest panel, filled pale yellow with a gold border, is titled in bold navy ``Phase 1: Monitoring,'' with the grey subtitle ``Daily diary intake, structured extraction, and change detection.'' At its far left is a solid black person silhouette labeled in bold ``User.'' To its right, a speech-bubble icon with three dots sits above a grey double-headed horizontal arrow, labeled ``Diary Intake Interaction'' with the smaller text ``(Morning / Evening)'' below. This arrow leads to a row of four white boxes with gold borders, linked left to right by small black triangular arrowheads. Each box has a bold title, a list of dashed items, and an icon in its top-right corner. The first box, marked with the OpenAI logo, is titled ``Diary Intake Conversation'' and lists ``Collect diary content'' and ``Follow-up questions.'' The second box, also marked with the OpenAI logo, is titled ``Text to JSON Converter'' and lists ``Text to JSON'' and ``Key information extraction,'' followed by the smaller parenthetical ``(sleep timing, mood, etc.)'' The third box, marked with the gear icon, is titled ``Baseline Creator/Updater'' and lists ``Create / update baseline'' and ``Maintain history.'' A black double-headed vertical arrow connects it to a database-cylinder icon below, labeled ``Update User Profile \& Baseline.'' The fourth box, also marked with the gear icon, is titled ``Change Detector \& Selector'' and lists ``Detect notable changes'' and ``Select topics for reflection.'' It is likewise connected by a double-headed vertical arrow to a database-cylinder icon labeled ``Update User History.'' The second panel, taller than the others and filled pale green with a green border, is titled in bold navy ``Phase 2: Reflection.'' At its top is a light-green box with a dark green border, marked with the OpenAI logo and containing the bold text ``Reflection Conversation.'' Below it is the bold green heading ``Reflection Decision Tree,'' followed by four numbered decision nodes that zigzag down the panel. The tree uses color to separate its parts. The decision nodes, which are the prompts the system tries in sequence, are white boxes with green borders and bold text. The branch labels, which give the user's possible responses, are written in red. The outcome boxes, which end each branch, are white with grey borders and regular text. Node ``1. Invite an initial interpretation'' branches ``Not sure'' down to node 2 and ``User Knows'' to an outcome reading ``Validate and explore + education.'' Node ``2. Prompt recall of the previous day'' branches ``User Knows'' to the same outcome and ``Not sure'' down to node 3. Node ``3. Introduce cues from the evening diary'' branches ``Not sure'' down to node 4 and ``User Knows'' to the same outcome. Node ``4. Assess whether it is recurring'' branches ``Yes'' to an outcome reading ``Recommend consult specialist'' and ``No'' to ``Validate and explore + education.'' The third panel, filled pale blue with a blue border, is titled in bold navy ``Phase 3: Education,'' with the grey subtitle ``Deliver personalized sleep education.'' It contains a light-blue box with a dark blue border, marked with the OpenAI logo, with the bold title ``Education Conversation'' and the text ``Based on user context (reflection) pick and relay education from the material provided by specialists.'' Together, the three color-coded phases illustrate the initial prototype's pipeline, which combines LLM-powered conversation and extraction with rule-based baseline construction and change detection. Morning and evening diaries collected through conversation are converted into structured records, compared against a personal baseline, and screened for notable changes to discuss. For a selected change, reflection escalates through up to four prompts until the user can account for it. A change the user can explain leads to validation, exploration, and education, while a change that remains unexplained and is recurring leads to a recommendation to consult a specialist. Education then relays specialist-provided material based on what reflection surfaced.}
\label{fig:stage2-architecture}
\end{figure*}

Building on the Stage~1 design findings, we implemented an initial prototype of \dream. The system was deployed as an Amazon Alexa skill, with an AWS Lambda skill handler and middleware and a backend hosted on Microsoft Azure, using the dual-API architecture introduced by Mahmood et al.~\cite{mahmood2025user} to accommodate LLM response latency. OpenAI's GPT-5.2 supported conversational diary intake, structured extraction, reflection, and response generation, while backend rules handled sleep-measure calculation, baseline construction, change detection, and change selection. Structured diary records and interaction histories were stored in separate Google Sheets for each participant. The system comprised three modules corresponding to the workflow established in Stage~1: \textit{monitoring}, \textit{reflection}, and \textit{education}. Figure~\ref{fig:stage2-architecture} provides an overview of the system architecture and the flow of information across these modules. Additional implementation details, including the prompting strategy and module-specific LLM handlers, are provided in Appendix~\ref{app:implementation}.

\begin{itemize}[leftmargin=*]
    \item \textbf{\highlight{lightyellow}{Monitoring}:} The monitoring module combined conversational intake with backend computation to process daily sleep records through the following steps:
    \begin{enumerate}[leftmargin=*]
        \item \textbf{Conversational Diary Intake.} The system conducted conversational sleep diaries every morning and evening (see Appendix~\ref{app:sleep-diary-questions} for the complete question lists). Once completed, these dialogues were converted into structured longitudinal records containing sleep measures and daily context.
        \item \textbf{Metric Derivation and Baseline Maintenance.} The backend derived TIB, TST, and SE from the morning diary and maintained a moving personal baseline for TIB, TST, SOL, WASO, SE, NWAK, and sleep quality. During the technology probe, the first three days established the initial baseline; subsequent observations were incorporated into the accumulated record, updating the mean and standard deviation of each measure.
        \item \textbf{Change Detection (Flagging).} Each completed morning record was evaluated using the specialist-informed thresholds developed in Stage~1 (Table~\ref{tab:initial-flags}), combining deviations from the personal baseline with absolute sleep criteria. We use \textit{flag} to denote a detected change meeting one of these criteria, generating \textit{soft personalized flags} for moderate deviations, \textit{hard personalized flags} for larger deviations, and \textit{literature-informed flags} for absolute threshold crossings.
        \item \textbf{Change Selection.} When multiple measures were flagged, one was selected for reflection according to the specialist-informed priority established in Stage~1. If no measure was flagged, the interaction ended after the morning diary.
    \end{enumerate}

    \item \textbf{\highlight{lightgreen}{Reflection}:} The reflection module followed the specialist-informed decision tree developed in Stage~1, using the selected sleep change, the participant's recent sleep pattern, and relevant context from the preceding evening diary. Reflection progressed to more directed support only when the user remained uncertain:
    \begin{enumerate}[leftmargin=*]
        \item \textbf{Invite an initial interpretation.} Present the selected change and ask what the user thinks may have contributed without first suggesting an explanation.
        \item \textbf{Prompt recall.} If the user is unsure, ask whether anything differed in the previous day, evening, or routine.
        \item \textbf{Introduce contextual cues.} If uncertainty remains, introduce one or two potentially relevant details from the preceding evening diary.
        \item \textbf{Assess frequency.} If no \textit{actionable contributor} emerges, ask whether the change appears isolated or recurring.
    \end{enumerate}

    \item \textbf{\highlight{lightblue}{Education}:} Following reflection, the education module used the selected sleep change and the user's interpretation to deliver tailored guidance. The system provided a brief explanation and one or two actionable strategies drawn from the specialist-provided sleep-hygiene material. In this initial prototype, every selected flag initiated reflection, and every completed reflection automatically proceeded to education.
\end{itemize}

\begin{table*}[t]
\caption{Flagging thresholds used in the initial prototype, informed by prior sleep literature and refined with sleep specialists.}
\label{tab:initial-flags}
\footnotesize
\renewcommand{\arraystretch}{1.15}
\setlength{\tabcolsep}{5pt}
\centering
\begin{tabular}{@{}p{2.2cm}p{4.0cm}p{4.0cm}p{4.0cm}@{}}
\toprule
\textbf{Measure} &
\textbf{Soft Personalized Flag} &
\textbf{Hard Personalized Flag} &
\textbf{Literature-Informed Flag} \\
\midrule
TST &
15--24\% below baseline mean &
$\geq 25\%$ below baseline mean &
$<6$ hours \\
SOL &
10--14 min above baseline mean &
$\geq 15$ min above baseline mean &
$>30$ min \\
WASO &
10--14 min above baseline mean &
$\geq 15$ min above baseline mean &
$>30$ min \\
TIB &
15--30 min above baseline mean &
$>30$ min above baseline mean &
TIB $>$ TST $+$ 45 min \\
SE &
5--7 percentage points below baseline mean &
$\geq 8$ percentage points below baseline mean &
$<85\%$ \\
NWAK &
2 awakenings above baseline mean &
$\geq 3$ awakenings above baseline mean &
$\geq 5$ awakenings \\
Sleep quality &
$>1\,SD$ below baseline mean &
--- &
--- \\
\bottomrule
\end{tabular}
\end{table*}

\section{Stage 3: Technology Probe and Formative Review}
\label{sec:stage3}

We deployed the initial prototype as a technology probe for 10 days in participants' homes to examine repeated use and generate real interactions for specialist review. The deployment was formative: its purpose was to identify where the initial monitoring, reflection, and education decisions required refinement before a longer field study.

\subsection{Participants and Procedure}
We recruited 11 undergraduate and graduate students through university mailing lists and campus flyers (ages 21--27, $M=23.09$, $SD=2.0$; eight women, three men). We refer to technology-probe participants as \texttt{TP-PXX}, where \texttt{XX} denotes the participant ID. The study was approved by the university's Institutional Review Board (IRB), and all participants provided informed consent. Participants were compensated \$30 for completing the study. Participants received an Echo Dot and completed a practice interaction during an in-person setup session. We configured personalized morning and evening diary windows around each participant's usual sleep and wake schedule; during each two-hour window, the system prompted participants to complete the corresponding diary and sent reminders until it was completed or the window ended. Days~1--3 initialized the personal baseline without reflection or education. During Days~4--10, a selected flag led to reflection and education; otherwise, the interaction ended after the diary. Participants completed post-study questionnaires and a semi-structured interview.

\subsection{Evaluation and Analysis}
We examined participant experiences, system behavior, and specialist assessments to evaluate the initial prototype and inform subsequent refinement.

\subsubsection{Participant Interviews}
Post-study interviews examined participants' experiences with repeated diary use, reflection, education, and interaction burden. The interview sessions were audio-recorded, transcribed, and iteratively analyzed to identify recurring themes. Several participants questioned the value of extended interactions when a detected change did not feel meaningful enough to warrant further discussion or was already sufficiently explained by a clear circumstance. \texttt{TP-P03}, for example, explained that the reason for her irregular sleep was \myc{\textit{``kind of obvious for me,''}} making further reflection feel \myc{\textit{``a bit unnecessary.''}} \texttt{TP-P07} similarly noted that \myc{\textit{``eventually I do think it got repetitive''}} when similar routines continued to produce similar interactions. These accounts suggested that detecting a change alone was not sufficient reason to initiate or continue reflection, particularly when the change was minor, expected, or already understood.

\subsubsection{Researcher Checks}
Two researchers reviewed all 49 flagged interactions to verify diary extraction, derived sleep measures, flagging logic, and whether the interaction followed the intended \textit{Noticing}, \textit{Reasoning}, and \textit{Action} flow. Five interactions were independently reviewed by both researchers, with agreement on all checks (100\%). Given the small overlap, we report this agreement descriptively. At the interaction level, 47 of 49 interactions (95.9\%) passed all researcher checks; the remaining two contained extraction or interpretation errors that produced an incorrect sleep value and flag.

\subsubsection{Specialist Review}
The two behavioral sleep medicine specialists reviewed all 49 flagged morning interactions across four stages: \textbf{Flagging}, whether the detected change warranted follow-up; \textbf{Noticing}, whether the change was presented accurately and appropriately; \textbf{Reasoning}, whether follow-up questions and contextual cues were appropriate and supported by the available information; and \textbf{Action}, whether subsequent education and suggested actions followed appropriately from the sleep issue and user's account. Flagging was coded as correct or incorrect, while noticing, reasoning, and action were rated for accuracy, appropriateness, relevance, and specificity on five-point scales.

Five interactions were independently reviewed by both specialists. They agreed on 4 of 5 flagging judgments (80.0\%). Across 60 paired stage-level ratings, 29 (48.3\%) matched exactly and 48 (80.0\%) differed by no more than one scale point. Given the small overlap, we report agreement descriptively.

Specialists judged 22 of the 49 flagged interactions (44.9\%) as appropriate and 27 (55.1\%) as inappropriate. Among rated interactions, \textit{Noticing} ($M=4.81$--$4.92$) and \textit{Reasoning} ($M=4.50$--$4.83$) were rated highly across the four criteria, whereas \textit{Action} showed greater variation ($M=3.64$--$4.67$). These results suggest that the prototype generally followed the intended workflow, but the system did not yet reliably determine when reflection was warranted. Full criterion-level ratings are reported in Appendix~\ref{app:specialist-evaluation}.

Specialist feedback during reviews illuminated why certain flags and interventions fell short. For instance, reviewing a minor deviation in time in bed, specialist \textit{S1} noted, \myc{\textit{``within that range there's going to be some variability,''}} while \textit{S2} emphasized that \myc{\textit{``if this happens every once in a while, it's really not an issue.''}} Furthermore, when the prototype moved too quickly from a user's ambiguous response to a causal conclusion, \textit{S2} observed, \myc{\textit{``I feel like there's an assumption here by the agent.''}} Finally, regarding the system's tendency to always deliver educational tips, \textit{S2} cautioned against tending to \myc{\textit{``over-inform or over-intervene''}} rather than \myc{\textit{``connecting it to what's actually happening.''}} These insights revealed that while the prototype implemented its conversational rules smoothly, its initial logic failed to capture ordinary variability, appropriate timing, or the conditionality needed for guidance.

\subsection{Design Refinements}
Based on participant experiences, researcher checks, and specialist reviews, we synthesized four key refinements for the refined system:

\begin{itemize}[leftmargin=*]
    \item \textbf{Monitoring: Separate detection from reflection eligibility.} Not every detected deviation warranted discussion. Informed by participant feedback questioning the value of discussing minor or expected changes, alongside specialist insights on ordinary variability and recurrence, we separated detecting a change from deciding whether it provided a sufficient basis for reflection.
    
    \item \textbf{Monitoring: Stabilize the longitudinal reference.} To prevent unusual values or single outlier nights from immediately skewing baseline comparisons, discussions with the specialists highlighted the need to stabilize the reference data. We therefore revised how and when observations entered the longitudinal reference and screened individual measures independently.
    
    \item \textbf{Reflection: Clarify before interpreting.} Prompted by specialist warnings against agent assumptions, we revised the reflection flow to clarify ambiguous or inconsistent user explanations before drawing connections or introducing contextual cues.
    
    \item \textbf{Education: Make guidance conditional on reflection.} Addressing participant fatigue with repetitive prompts when a change was already understood, and specialist concerns about over-intervention, we made education conditional on what was established through reflection, allowing interactions to end with a brief acknowledgment or without guidance when further support was unnecessary.
    
\end{itemize}

\section{Stage 4: Refining \dream}
\label{sec:final-system}

\begin{figure*}[t]
\centering
\includegraphics[width=\textwidth]{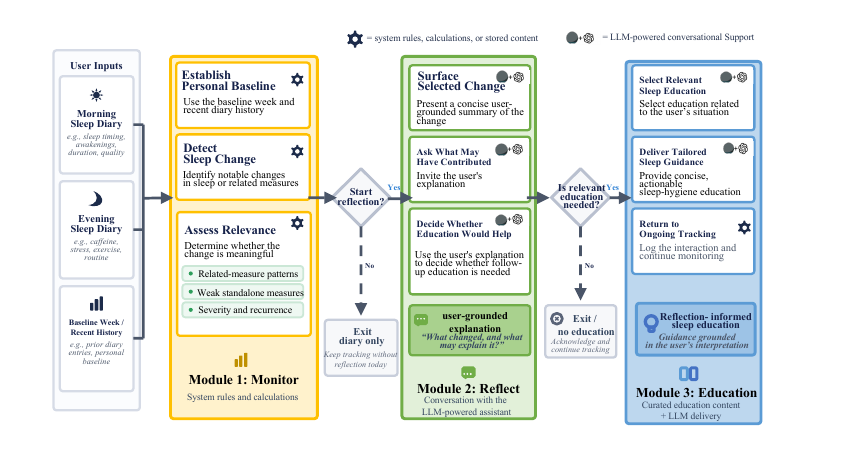}
\caption{The refined \dream pipeline. Morning and evening diaries produce structured records; detected changes pass through a separate eligibility check before reflection begins; and reflection determines whether education follows or the interaction ends with a brief acknowledgment. Gear icons mark steps implemented as backend rules; assistant icons mark LLM-powered conversation.}
\Description{This figure shows a left-to-right flowchart made up of a user-input column, three module panels, and two decision diamonds between the modules. The three modules use the same color scheme as the other figures in the paper: yellow for Monitor, green for Reflect, and blue for Education. The white step boxes inside each module have borders in that module's color, marking which module each step belongs to. Elements outside the modules, including the user-input column, the decision diamonds, and the exit boxes, are drawn in grey. A legend along the top defines two icon types. A dark navy gear means ``= system rules, calculations, or stored content.'' A dark grey smart-speaker icon joined by a plus sign to the OpenAI logo means ``= LLM-powered conversational Support.'' At the far left, a panel with a light-grey border is headed in bold ``User Inputs'' and contains three stacked boxes, each with an icon, a bold title, and grey italic examples. The first shows a sun icon and ``Morning Sleep Diary,'' with ``e.g., sleep timing, awakenings, duration, quality.'' The second shows a crescent-moon icon and ``Evening Sleep Diary,'' with ``e.g., caffeine, stress, exercise, routine.'' The third shows a bar-chart icon and ``Baseline Week / Recent History,'' with ``e.g., prior diary entries, personal baseline.'' Lines from the three boxes join into a bracket that feeds a single thick dark-grey arrow into the first module. The first module, filled pale yellow with a yellow border, contains three stacked white boxes, each marked with the gear icon. The first is titled ``Establish Personal Baseline'' with the text ``Use the baseline week and recent diary history.'' The second is titled ``Detect Sleep Change'' with ``Identify notable changes in sleep or related measures.'' The third is titled ``Assess Relevance'' with ``Determine whether the change is meaningful,'' followed by three light-green pill-shaped items with green bullet points: ``Related-measure patterns,'' ``Weak standalone measures,'' and ``Severity and recurrence.'' At the bottom of the module, a small gold bar-chart icon sits above the bold label ``Module 1: Monitor'' and the text ``System rules and calculations.'' An arrow leaves the right side of Module 1 and enters a grey-outlined diamond labeled ``Start reflection?'' A ``Yes'' branch, labeled in blue, continues right into the second module. A ``No'' branch runs down as a dashed arrow to a grey-outlined box reading ``Exit diary only,'' with the italic text ``Keep tracking without reflection today.'' The second module, filled pale green with a green border, contains three stacked white boxes, each marked with the smart-speaker and OpenAI logo icon. The first is titled ``Surface Selected Change'' with ``Present a concise user-grounded summary of the change.'' The second is titled ``Ask What May Have Contributed'' with ``Invite the user's explanation.'' The third is titled ``Decide Whether Education Would Help'' with ``Use the user's explanation to decide whether follow-up education is needed.'' Below these is a box filled a deeper green, containing a green speech-bubble icon, the bold text ``user-grounded explanation,'' and the italic question ``What changed, and what may explain it?'' set in quotation marks. At the bottom of the module, a green speech-bubble icon sits above the bold label ``Module 2: Reflect'' and the text ``Conversation with the LLM-powered assistant.'' An arrow leaves the right side of Module 2 and enters a second grey-outlined diamond labeled ``Is relevant education needed?'' A blue ``Yes'' branch continues right into the third module. A ``No'' branch runs down as a dashed arrow to a grey-outlined box with a circled-X icon, reading ``Exit / no education,'' with the italic text ``Acknowledge and continue tracking.'' The third module, filled pale blue with a blue border, contains three stacked white boxes. The first, marked with the smart-speaker and OpenAI logo icon, is titled ``Select Relevant Sleep Education'' with ``Select education related to the user's situation.'' The second, also marked with that icon, is titled ``Deliver Tailored Sleep Guidance'' with ``Provide concise, actionable sleep-hygiene education.'' The third, marked with the gear icon, is titled ``Return to Ongoing Tracking'' with ``Log the interaction and continue monitoring'' in grey text. Below these is a box filled a deeper blue, containing a blue lightbulb icon, the bold text ``Reflection-informed sleep education,'' and the italic text ``Guidance grounded in the user's interpretation.'' At the bottom of the module, a blue icon of two rounded rectangles side by side sits above the bold label ``Module 3: Education'' and the text ``Curated education content + LLM delivery.'' In Modules 2 and 3, the deeper-filled box below the step boxes highlights what that module produces. Together, the three color-coded modules illustrate the refined DREAM pipeline. Morning and evening diaries, along with the baseline week and recent history, feed rule-based monitoring, which establishes a personal baseline, detects sleep changes, and assesses whether a change is meaningful before any conversation begins. If reflection is not started, the day ends with diary tracking only. Otherwise, LLM-powered reflection surfaces the change, invites the user's own explanation, and uses that explanation to decide whether education would help. When it would, the system selects and delivers tailored guidance grounded in the user's interpretation. When it would not, the system acknowledges the user and continues tracking. In either case, the interaction returns to ongoing monitoring.}
\label{fig:dream-pipeline}
\end{figure*}

The refined system retained the initial platform, conversational diaries, and three-module structure, but restructured how detected sleep changes led to reflection and how reflection shaped subsequent education. Fig.~\ref{fig:dream-pipeline} summarizes the resulting workflow, and Table~\ref{tab:example-episodes} illustrates two interactions from the field study.

\subsection{Refinements to the \highlight{lightyellow}{Monitoring} Module}

\subsubsection{Stable Baseline Updating}
To prevent unusual values or single outlier nights from immediately skewing baseline comparisons, Week~1 established the initial baseline pool. During each subsequent week, new observations were evaluated against the cumulative record available before that week began and were added only after the week was complete. Before being added, morning sleep measures were screened independently so that an unusual value in one measure did not exclude otherwise usable values from the same diary.

To filter out anomalies while respecting the natural physiological variability of sleep, we established specialist-informed screening equations. Specifically, a morning value was excluded from the baseline pool when its absolute deviation from the baseline median exceeded $\max(2\times\mathrm{MAD},\text{ metric-specific floor})$. The metric-specific floors and thresholds were informed by specialist discussions, aligning with the deviation magnitudes used for hard personalized candidate detection: $0.25\times$ the median for TST, 15 minutes for SOL and WASO, 8 percentage points for SE, three awakenings for NWAK, and 30 minutes for TIB. Evening stress, mood, pain, and fatigue were screened using $2\times\mathrm{MAD}$. These screening rules determined which observations entered the longitudinal baseline and were separate from the rules determining whether reflection occurred.

\subsubsection{Candidate Detection and Reflection Eligibility}
Informed by specialist insights on ordinary variability and participant feedback regarding interaction burden, the refined system separated detecting a change from deciding whether it warranted reflection. Candidate detection retained the soft personalized, hard personalized, and literature-informed thresholds used in the initial prototype (Table~\ref{tab:initial-flags}). We use \textit{candidate} to refer to a detected change before the eligibility decision. Unlike a flag in the initial prototype, a candidate did not automatically initiate reflection.

A separate backend gate determined reflection eligibility based on severity, recurrence, and relationships among sleep measures:

\begin{enumerate}[leftmargin=*]

\item \textbf{Related-measure patterns.}
Certain combinations were eligible immediately: \textit{fragmentation}, defined by at least two of elevated WASO, elevated NWAK, and reduced SE; \textit{efficiency}, defined by elevated TIB with reduced SE; and \textit{sleep opportunity}, defined by reduced TST with elevated TIB.

\item \textbf{Weak standalone changes.}
Standalone TIB was suppressed when SE was at least 85\% and no higher-priority candidate was present. Standalone NWAK with WASO below 30 minutes, and standalone SOL or WASO at or below 30 minutes, required a second occurrence within seven days.

\item \textbf{Strong standalone changes.}
Remaining hard personalized and literature-informed candidates were eligible immediately.

\item \textbf{Recurring soft changes.}
Remaining soft personalized candidates became eligible after a second occurrence of the same measure within seven days.

\end{enumerate}

Candidates that were not yet eligible remained in the longitudinal record for recurrence checks, while the interaction ended after the diary. When multiple candidates were eligible, related-measure patterns were prioritized when present; otherwise, stronger candidates were prioritized before softer candidates, followed by the measure-level ordering used in the initial prototype.

\subsection{Refinements to the \highlight{lightgreen}{Reflection} Module}
Prompted by specialist warnings against premature agent assumptions, we introduced two refinements to how reflection proceeded:

\begin{enumerate}[leftmargin=*]

\item \textbf{Clarify before interpreting.}
Reflection began by presenting the selected change neutrally and asking what the user thought may have contributed. When the user's explanation was ambiguous or inconsistent with recorded information, the LLM asked clarification questions rather than drawing a connection itself. Recorded context was introduced only when the user remained uncertain and was framed as a possible cue rather than an inferred cause.

\item \textbf{Allow early exit for one-time explanations.}
When a clear one-time circumstance, such as travel, an examination, an appointment, or an unusual obligation, sufficiently explained the selected change, the system briefly acknowledged the explanation and ended the interaction without further probing.
\end{enumerate}

\subsection{Refinements to the \highlight{lightblue}{Education} Module}
Addressing participant fatigue with repetitive prompts and specialist concerns over over-intervention, education was no longer an automatic consequence of reflection. When reflection identified or clarified an \textit{actionable contributor} that could reasonably be connected to the selected sleep issue, the system provided a brief explanation of that connection and one or two actions drawn from the specialist-provided sleep-hygiene material.

Reflection could therefore end in three ways: (1) an actionable contributor led to education; (2) a clear one-time explanation led to acknowledgment and exit; or (3) when no actionable contributor emerged, the interaction ended without guidance. In the refined system, reflection therefore determined both what education should address and whether education was warranted at all.

\begin{table*}[t]
\caption{Two \dream interactions from the participants illustrating how guided reflection could lead to different outcomes. Left: reflection identified a potential cause and led to education. Right: reflection indicated that the change was a one-time occurrence and ended without education.}
\label{tab:example-episodes}
\footnotesize
\renewcommand{\arraystretch}{1.15}
\setlength{\tabcolsep}{4pt}
\centering
\begin{tabularx}{\textwidth}{@{}p{2.5cm}>{\raggedright\arraybackslash}X>{\raggedright\arraybackslash}X@{}}
\toprule
\textbf{Stage} &
\textbf{Education Interaction} &
\textbf{Exit Interaction} \\
\midrule

\textbf{\highlight{lightyellow}{Monitoring}}\newline
\textbf{Selected Change}
&
The participant was awake for about 30 minutes during the night, compared with a personal baseline of about 4 minutes.
&
The participant spent about six and a half hours in bed, slept about five and a quarter hours, and had awakened about two hours earlier than planned.
\\
\addlinespace[4pt]

\textbf{\highlight{lightgreen}{Reflection}}\newline
\textbf{Noticing}
&
\textit{System: ``I noticed you were awake during the night for about thirty minutes last night, and you are usually awake only around four minutes. What do you think might have contributed to that?''}

\textit{User: ``I had to go to the bathroom in the middle of my sleep.''}
&
\textit{System: ``I noticed you spent about six and a half hours in bed last night, but you only slept about five and a quarter hours. What do you think might have contributed to that last night?''}

\textit{User: ``I had something scheduled in the morning.''}
\\
\addlinespace[4pt]

\textbf{\highlight{lightgreen}{Reflection}}\newline
\textbf{Guided Reflection}
&
The system followed the participant's explanation to identify what may have contributed to the nighttime awakening:

\textit{System: ``What do you think led to needing to use the bathroom during the night, like drinking more later than usual, or anything else different yesterday evening?''}

\textit{User: ``I drank a lot of water.''}
&
The system clarified what had prompted the earlier awakening and whether it reflected a recurring pattern:

\textit{System: ``Was there something specific scheduled for this morning?''}

\textit{User: ``A CVS appointment.''}

\textit{System: ``Does that feel like more of a one time thing, or something that has been happening a lot lately?''}

\textit{User: ``It's a one time thing.''}
\\
\addlinespace[4pt]

\textbf{Outcome}
&
\textbf{\highlight{lightblue}{Education}}

\textit{System: ``Drinking a lot in the evening can sometimes lead to waking up to use the bathroom. One thing you could try is having most of your water earlier in the day and going lighter on fluids close to bedtime.''}

The participant agreed to reduce fluid intake closer to bedtime.
&
\textbf{Acknowledgment and Exit}

\textit{System: ``Yeah, that tracks. A one time morning appointment can nudge your body into waking earlier. Sounds like that explains last night.''}

The interaction ended without education.
\\

\bottomrule
\end{tabularx}
\end{table*}


\section{Stage 5: Deploying \dream in the Field}
\label{sec:field-study}

\begin{figure*}[t!]
\centering
\includegraphics[width=\textwidth]{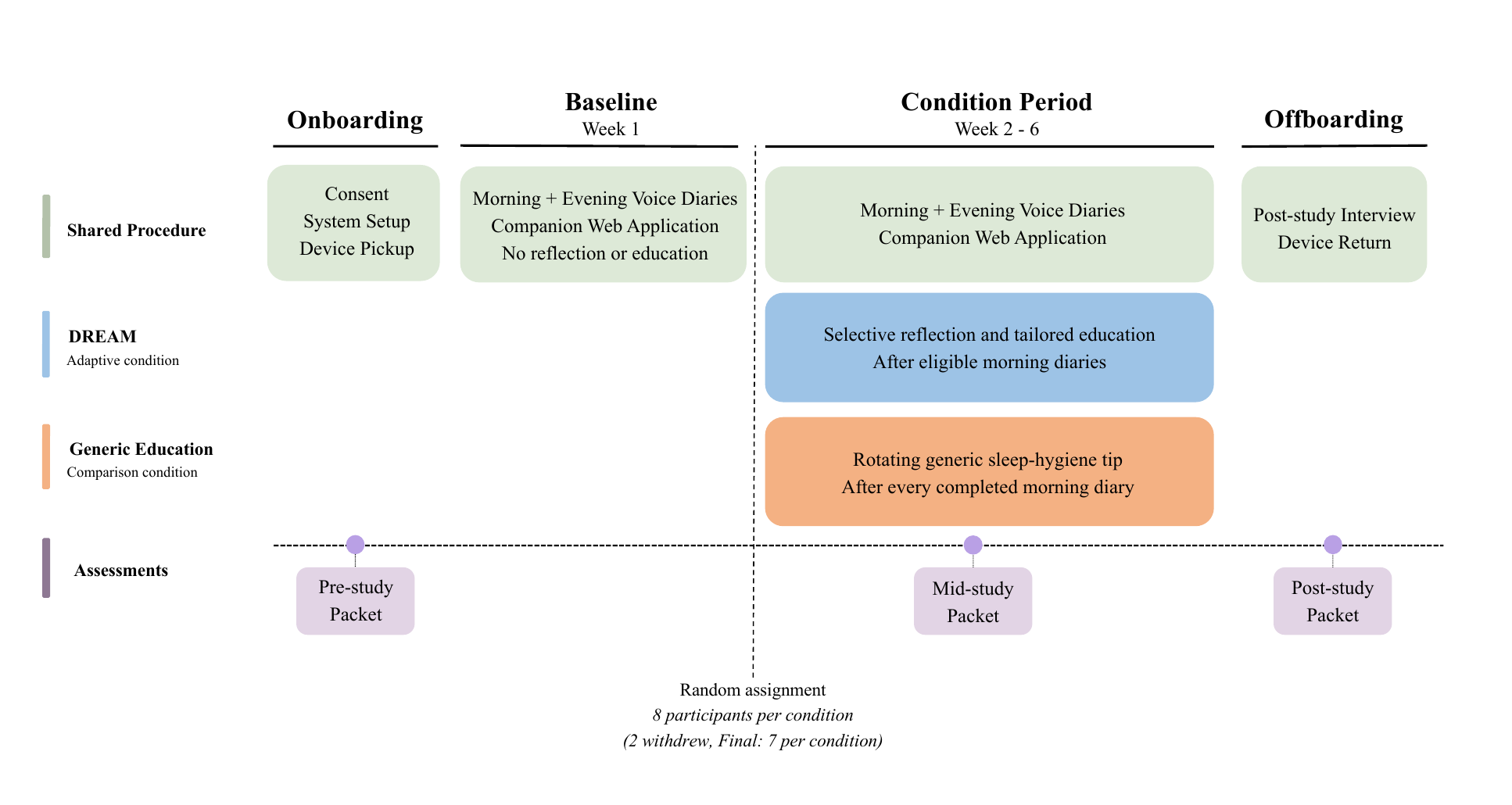}
\caption{Procedure of the six-week field study. All participants completed the same morning and evening conversational diaries during a shared one-week baseline. During Weeks~2--6, participants in the \dream condition received selectively initiated reflection and tailored education, whereas participants in the Generic Education (control) condition received one rotating generic sleep-hygiene tip after each completed morning diary. Questionnaire packets were completed before deployment, midway through condition-specific use, and after Week~6.}
\Description{This figure shows a study timeline laid out as a grid, with four phases as columns and four tracks as rows. The rows are color-coded, and a short vertical bar next to each row label matches the color of the boxes in that row. Pale green marks the shared procedure that all participants followed. Blue marks the DREAM condition, orange marks the Generic Education condition, and purple marks the assessments. Across the top, four bold phase headings each sit above a black horizontal line: ``Onboarding,'' ``Baseline'' with ``Week 1'' beneath it, ``Condition Period'' with ``Week 2 - 6'' beneath it, and ``Offboarding.'' Down the left side, the four row labels are the bold ``Shared Procedure''; the bold ``DREAM'' with the smaller text ``Adaptive condition''; the bold ``Generic Education'' with the smaller text ``Comparison condition''; and the bold ``Assessments.'' In the Shared Procedure row, a pale green rounded box appears under each of the four phases. The Onboarding box reads ``Consent,'' ``System Setup,'' and ``Device Pickup.'' The Baseline box reads ``Morning + Evening Voice Diaries,'' ``Companion Web Application,'' and ``No reflection or education.'' The Condition Period box reads ``Morning + Evening Voice Diaries'' and ``Companion Web Application.'' The Offboarding box reads ``Post-study Interview'' and ``Device Return.'' The DREAM and Generic Education rows are empty except under the Condition Period, where each has one wide rounded box spanning the phase. The blue DREAM box reads ``Selective reflection and tailored education'' and ``After eligible morning diaries.'' The orange Generic Education box, directly below it, reads ``Rotating generic sleep-hygiene tip'' and ``After every completed morning diary.'' In the Assessments row, a horizontal dashed line runs the full width of the timeline. Three purple circular markers sit on this line, each connected by a short dotted line to a pale purple rounded box below. The first marker, under Onboarding, leads to ``Pre-study Packet.'' The second, midway through the Condition Period, leads to ``Mid-study Packet.'' The third, under Offboarding, leads to ``Post-study Packet.'' A vertical dashed line separates the Baseline and Condition Period columns, running from the top of the grid down through all four rows. At its lower end is the label ``Random assignment,'' followed by the italic text ``8 participants per condition'' and ``(2 withdrew, Final: 7 per condition).'' Together, the four rows illustrate the six-week field study. All participants shared onboarding, a one-week baseline of morning and evening voice diaries without reflection or education, continued diary use throughout, and a common offboarding. After random assignment at the end of Week 1, the two conditions diverged only during Weeks 2 to 6. DREAM participants received selective reflection and tailored education after eligible morning diaries, while Generic Education participants received a rotating generic sleep-hygiene tip after every completed morning diary. Questionnaire packets were completed at the start, midway through the condition period, and at the end.}
\label{fig:study-timeline}
\end{figure*}

We conducted a six-week, between-subjects field study comparing \dream with a Generic Education control condition, consisting of a shared one-week baseline followed by five weeks of condition-specific use (Fig.~\ref{fig:study-timeline}).

\subsection{Study Design and Conditions}
\label{sec:study-design}
Both conditions used the same voice diaries, reminder schedule, companion web application, and specialist-provided sleep-hygiene material. The companion web application was included to approximate common consumer sleep-tracking tools, where users can review their tracked sleep data alongside educational support~\cite{arroyo2022sleepapps}. 

\begin{figure*}[t]
    \centering
    \includegraphics[width=\textwidth]{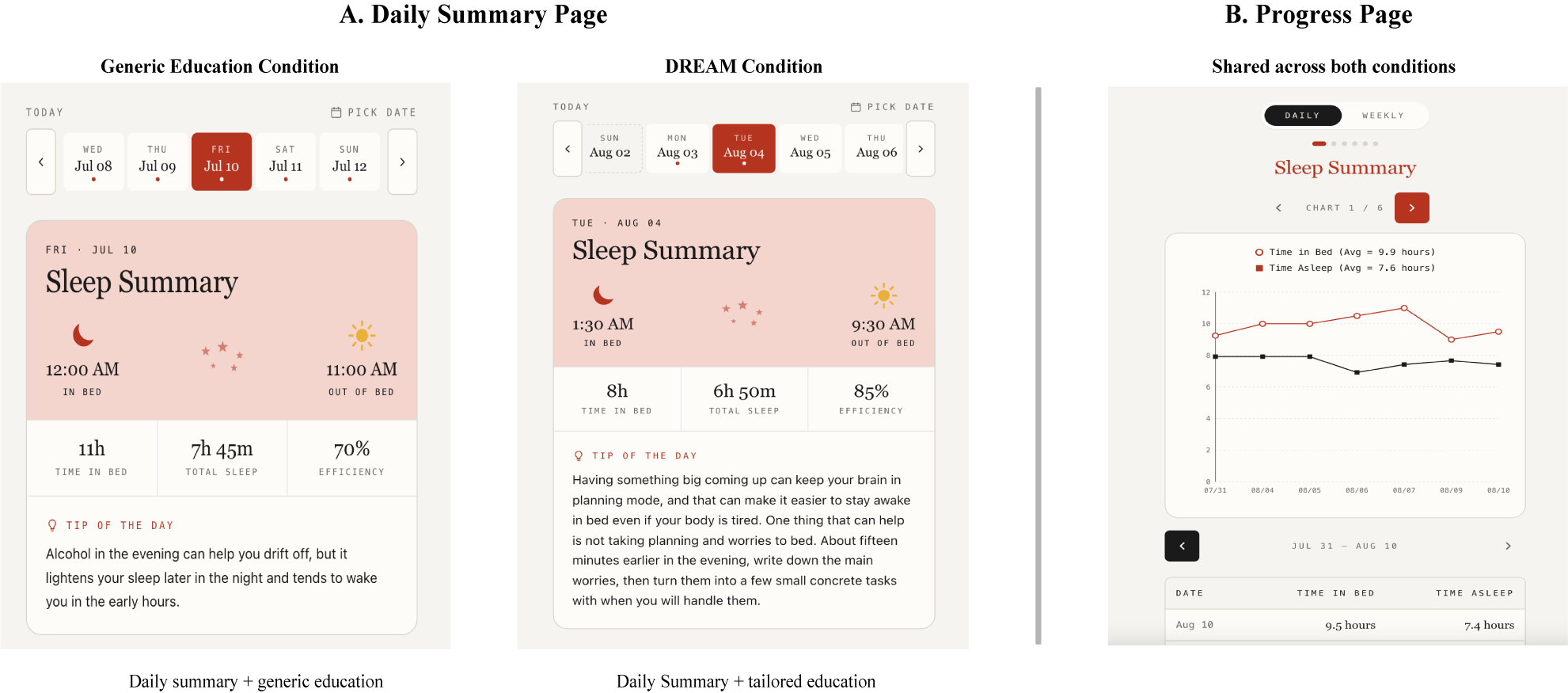}
    \caption{The companion web application used during the field study.
    (A) The Daily Summary page presented nightly sleep measures alongside education. Participants in the Generic Education condition received generic education independent of their sleep data, whereas \dream participants received tailored education when reflection led to guidance.
    (B) The Progress page was identical across conditions and allowed participants to review longitudinal changes in their sleep measures. 
    }
    \Description{This figure shows three screenshots of a web application arranged side by side under two bold headings. The heading ``A. Daily Summary Page'' spans the left two screenshots, and the heading ``B. Progress Page'' sits above the right screenshot. A thin vertical grey line separates the two groups. Above each screenshot is a bold subheading: ``Generic Education Condition'' on the left, ``DREAM Condition'' in the middle, and ``Shared across both conditions'' on the right. Below the left and middle screenshots are the labels ``Daily summary + generic education'' and ``Daily Summary + tailored education.'' All three screenshots use the application's own color palette on a light beige background, not the module colors used in the paper's other figures. Dark red marks the currently selected or active element, such as the chosen date and the active navigation button, and is also used for headings and chart data. A pale pink panel highlights the bedtime and wake time at the top of each sleep summary card. The two Daily Summary screenshots share an identical layout. At the top, the small spaced capitals ``TODAY'' appear on the left and a calendar icon with ``PICK DATE'' on the right. Below is a horizontal strip of five date tiles between left and right arrow buttons. Each tile shows a weekday abbreviation above a date, and the selected tile is filled dark red with white text. Small red dots appear beneath some of the dates. In the Generic Education screenshot, the tiles read ``WED Jul 08,'' ``THU Jul 09,'' ``FRI Jul 10'' (selected), ``SAT Jul 11,'' and ``SUN Jul 12,'' and every tile has a dot. In the DREAM screenshot, the tiles read ``SUN Aug 02,'' ``MON Aug 03,'' ``TUE Aug 04'' (selected), ``WED Aug 05,'' and ``THU Aug 06.'' Here ``SUN Aug 02'' has a dashed outline, and only ``MON Aug 03'' and the selected tile carry dots. Below the date strip is a card whose upper portion is pale pink. It shows the selected date in small capitals, followed by the large title ``Sleep Summary.'' Beneath the title, a red crescent-moon icon on the left sits above a bedtime labeled ``IN BED,'' a yellow sun icon on the right sits above a wake time labeled ``OUT OF BED,'' and a few small red stars sit between them. The lower portion of the card is white. It contains a row of three cells labeled ``TIME IN BED,'' ``TOTAL SLEEP,'' and ``EFFICIENCY,'' followed by a red lightbulb icon and the red heading ``TIP OF THE DAY'' above a paragraph of tip text. In the Generic Education screenshot, the card is dated ``FRI'' and ``JUL 10,'' shows ``12:00 AM'' in bed and ``11:00 AM'' out of bed, and lists ``11h,'' ``7h 45m,'' and ``70\%.'' Its tip reads ``Alcohol in the evening can help you drift off, but it lightens your sleep later in the night and tends to wake you in the early hours.'' In the DREAM screenshot, the card is dated ``TUE'' and ``AUG 04,'' shows ``1:30 AM'' in bed and ``9:30 AM'' out of bed, and lists ``8h,'' ``6h 50m,'' and ``85\%.'' Its longer tip reads ``Having something big coming up can keep your brain in planning mode, and that can make it easier to stay awake in bed even if your body is tired. One thing that can help is not taking planning and worries to bed. About fifteen minutes earlier in the evening, write down the main worries, then turn them into a few small concrete tasks with when you will handle them.'' The Progress Page screenshot begins with a two-option toggle in which ``DAILY'' is selected, shown as white text on a black pill, beside ``WEEKLY.'' Below the toggle is a row of six small page-indicator dots, the first drawn as a short red dash, followed by the red title ``Sleep Summary.'' Beneath the title are a left arrow, the text ``CHART 1 / 6,'' and a dark red square button with a right arrow. A white card below contains a line chart. Its legend reads ``Time in Bed (Avg = 9.9 hours),'' marked by a hollow red circle, and ``Time Asleep (Avg = 7.6 hours),'' marked by a filled square. The y-axis runs from 0 to 12 in steps of 2, and the x-axis is labeled ``07/31,'' ``08/04,'' ``08/05,'' ``08/06,'' ``08/07,'' ``08/09,'' and ``08/10.'' A red line with hollow circle markers, representing time in bed, stays between roughly 9 and 11 hours, peaking near 11 on 08/07 and dipping to about 9 on 08/09. A black line with filled square markers, representing time asleep, stays between roughly 7 and 8 hours. Below the chart are a black square button with a left arrow, the date range ``JUL 31 --- AUG 10,'' and a plain right arrow. At the bottom is a table with the column headers ``DATE,'' ``TIME IN BED,'' and ``TIME ASLEEP.'' Its first visible row reads ``Aug 10,'' ``9.5 hours,'' and ``7.4 hours,'' and the rest of the table is cut off. Together, the screenshots illustrate that both conditions used the same companion web application. On the Daily Summary page, participants in both conditions saw the same nightly sleep measures, and the conditions differed only in the tip of the day. The Generic Education tip is a general statement about sleep, while the DREAM tip addresses a specific situation, having something big coming up, and suggests a concrete action to take earlier in the evening. The Progress page, identical in both conditions, let participants review trends in their time in bed and time asleep over the preceding days.}
    \label{fig:companion-app}
\end{figure*}

In the \textbf{\dream} condition, each completed morning diary during Weeks~2--6 was processed by the refined system (Section~\ref{sec:final-system}): reflection began only when a candidate met the eligibility criteria, and education followed only when reflection established a relevant contributor. In the \textbf{Generic Education (control)} condition, participants received one tip after each completed morning diary, drawn from a fixed rotation of the same material and independent of their data, approximating non-personalized support in consumer sleep applications~\cite{arroyo2022sleepapps}. Control participants therefore received education every morning, whereas \dream participants received it only when reflection warranted it. The web application (Fig.~\ref{fig:companion-app}) showed daily and weekly views of TIB, TST, SE, sleep and wake times, SOL, and WASO in both conditions and displayed the day's education when provided.

\begin{figure}[t]
    \centering
    \includegraphics[width=\linewidth]{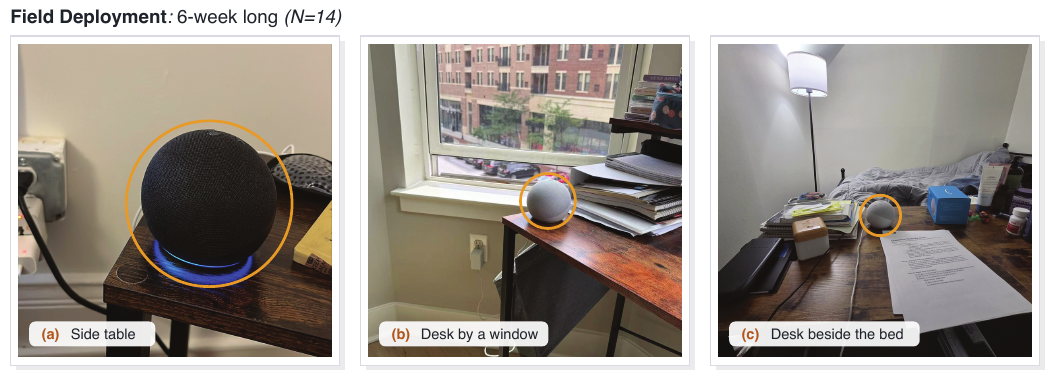}
    \caption{Participants placed the Amazon Echo Dot (circled) where it suited them during
    the six-week deployment, for instance: (a) a side table; (b) a desk beside a
    window; and (c) a desk next to the bed.}
    \Description{Three square photographs of an Amazon Echo Dot in participants'
    homes: on a side table with its light ring lit blue, on a wooden desk beside a
    window, and on a desk next to a bed with printed study materials.}
    \label{fig:deployment}
\end{figure}

\subsection{Participants and Procedure}
We recruited adults through university mailing lists. The study was approved by the university's Institutional Review Board (IRB), and all participants provided informed consent. Participants received \$50 for completing the study. Sixteen adults enrolled and were randomly assigned, eight per condition. Two participants withdrew during the study: one \dream participant because of travel and one control participant who chose not to continue. 

Our final sample comprised 14 adults (9 women, 5 men; age $M=26.14$, $SD=11.69$). Participants came from diverse educational and professional backgrounds and possessed advanced to native English fluency. On a 5-point scale where 1 indicated no prior experience and 5 indicated extensive experience, prior experience was moderate for general voice assistants ($M=2.64$, $SD=1.50$) and smart speakers ($M=2.36$, $SD=1.55$).

At onboarding, participants gave consent, completed a trial interaction, received an Echo Dot (Fig. \ref{fig:deployment}) and a web-application account, and had diary windows and reminders configured to their schedule. They completed the pre-study questionnaire packet after this session, before the study began. Week~1 established the baseline with diaries only. During Weeks~2--6, participants continued the diaries with condition-specific support, completed the packet again at mid-study, and after Week~6 completed the final packet, post-study measures, and a semi-structured interview. Participants are identified by condition and study ID throughout the paper, with \texttt{\dreamNospace-PXX} denoting the \dream condition and \texttt{CTRL-PXX} the control condition.

\subsection{Measures and Analysis}


\dream was designed to provide an interpretive layer between sleep tracking and subsequent guidance. We therefore expected \dream participants to show stronger outcomes related to personal sleep understanding, motivation, and system experience than participants receiving generic education. We additionally expected general sleep knowledge and self-efficacy to improve over the study period. Broader behavioral and sleep outcomes were examined as exploratory secondary outcomes because the primary focus of this work is on guided reflection leading to better understanding of sleep.

We used one-tailed between-condition tests for the five post-study measures directly related to the intended experience of \dream---willingness to use the system, disclosure comfort, personal sleep insight, sleep-data sensemaking, and motivation to improve sleep (full items in Appendix~\ref{app:poststudy-items})---and for three open-ended questions. Responses to these open-ended questions were de-identified and coded from 0 to 2 without condition labels (coding criteria and inter-rater agreement are detailed in Appendix~\ref{app:openended-coding}). Welch's $t$-tests were used for post-study measures meeting parametric assumptions and Wilcoxon rank-sum tests otherwise; coded open-ended responses were compared using Wilcoxon rank-sum tests. We report Cohen's $d$ for Welch's $t$-tests and rank-biserial correlations ($r_{rb}$) for Wilcoxon tests.

Repeated questionnaire measures of sleep knowledge and confidence (Sleep Self-Efficacy, Sleep Beliefs, and SHAPS knowledge) were analyzed longitudinally using mixed-effects models accounting for repeated observations within participants, with additional directional pre-to-post contrasts. Broader sleep and behavioral outcomes, including PSQI, SHAPS sleep-hygiene practices, nightly sleep measures, and routinely recorded behaviors, were summarized descriptively, with exploratory longitudinal analyses reported in Appendix~\ref{app:distal-outcomes}.

Additional details for specific measures are provided below.

\subsubsection{System Evaluation}

\paragraph{Researcher Checks.}
Two researchers reviewed 60 of the 113 \dream reflection interactions (53.1\%) using the Stage~3 researcher-check procedure, adapted to the refined workflow. They verified the intended \textit{Noticing}, \textit{Reasoning}, and \textit{Action} flow, including whether the interaction appropriately proceeded to education or ended without further guidance. Seven interactions were independently reviewed by both researchers to assess inter-rater agreement.

\paragraph{Specialist Review.}
The two behavioral sleep medicine specialists reviewed a separate sample of 60 of the 113 reflection interactions (53.1\%). Using the four-stage protocol from the Stage~3 technology probe (Section~\ref{sec:stage3}), they assessed whether the selected change provided an appropriate basis for reflection (\textit{Flagging}) and rated the accuracy, appropriateness, relevance, and specificity of \textit{Noticing}, \textit{Reasoning}, and \textit{Action}. Seven interactions were independently reviewed by both specialists to assess inter-rater agreement.

\subsubsection{System Use and Experience}
Following Week~6, participants completed the System Usability Scale (SUS)~\cite{brooke1996sus} and custom 5-point Likert measures of willingness to use the system (5 items; $\alpha=.93$) and disclosure comfort (4 items; $\alpha=.87$). System logs provided interaction timestamps, from which we calculated each participant's median morning interaction duration during Weeks~2--6 and summarized these durations descriptively by condition.

\subsubsection{Sleep Understanding and Action} \paragraph{Personal Sleep Understanding.} 
The post-study questionnaire assessed personal sleep insight (2 items; $\alpha=.82$) and sleep-data sensemaking (2 items; $\alpha=.71$); item scores were averaged within each measure. Participants also answered an open-ended question about what they learned about their own sleep, which was coded for personal specificity, insight depth, and relational reasoning. 

\paragraph{Sleep Knowledge.} Participants answered an open-ended question about what they learned about sleep more generally, which was coded for knowledge depth, specificity, practical relevance, and relational reasoning. Sleep-related knowledge and beliefs were also assessed at pre-, mid-, and post-study using the Sleep Beliefs Scale~\cite{adan2006sleepbeliefs} and the knowledge components of the Sleep Hygiene Awareness and Practice Scale (SHAPS)~\cite{lacks1986shaps}.

\paragraph{Motivation and Uptake.} Motivation to improve sleep was assessed with two 5-point Likert items ($\alpha=.83$) and averaged. Sleep Self-Efficacy~\cite{lacks1987sleepselfefficacy} was assessed at pre-, mid-, and post-study. Participants also answered an open-ended question about sleep-related strategies they learned or tried, which was coded for strategy specificity, behavioral uptake, and outcome reflection.

\subsubsection{Sleep and Behavioral Outcomes}
We examined broader sleep and behavioral outcomes to characterize changes over the six-week deployment. Overall sleep quality was assessed at pre-, mid-, and post-study using the Pittsburgh Sleep Quality Index (PSQI)~\cite{buysse1989psqi}. We additionally examined SHAPS sleep-hygiene practices, nightly sleep measures (TST, SOL, WASO, SE, sleep quality, and TIB), and routinely recorded sleep-related behaviors.

\subsubsection{Participant Perspectives}
We conducted semi-structured post-study interviews examining participants' experiences with the diary system, what they learned about their sleep, perceived relevance and burden, willingness to continue use, and sleep-related changes they considered or made. Interviews were audio-recorded, transcribed, and analyzed using thematic analysis; one author coded transcripts iteratively and developed preliminary themes, which were refined with the research team.

\subsection{Findings}

\subsubsection{System Evaluation}

\paragraph{Researcher Checks.}
The two researchers agreed on all checks for the seven independently reviewed interactions (100\%). Across the 60 reviewed interactions, \textit{Noticing} followed the intended flow in 58 interactions (96.7\%), while \textit{Reasoning} and \textit{Action} each did so in 55 interactions (91.7\%). Deviations included cases in which a participant identified a clear one-time explanation but the system continued the reflection rather than taking the intended acknowledgment-and-exit pathway. These deviations primarily added unnecessary conversational turns or probing rather than changing the substantive guidance provided. Overall, most reviewed interactions followed the intended refined workflow.

\paragraph{Specialist Review.}
For seven interactions independently reviewed by both specialists, they agreed on 6 of 7 flagging judgments (85.7\%). Among ratings available from both specialists, there were 20 paired ratings each for \textit{Noticing} and \textit{Reasoning} and 9 for \textit{Action}, because some interactions ended without education; 24 of 49 (49.0\%) matched exactly and 31 of 49 (63.3\%) differed by no more than one scale point. Given the small overlap, we report agreement descriptively.

Across the 60 reflection interactions reviewed by the specialists, 52 (86.7\%) were judged to provide an appropriate basis for follow-up. For these interactions, \textit{Noticing} received consistently high ratings ($M=4.88$--$4.98$), while \textit{Reasoning} ratings ranged from $M=4.35$--$4.46$. Eleven interactions ended without education, while 41 proceeded to education; \textit{Action} ratings among these 41 interactions ranged from $M=3.63$--$4.27$. Full criterion-level ratings, standard deviations, and the number of ratings contributing to each estimate are reported in Appendix~\ref{app:specialist-evaluation}.

The refined system more closely aligned with specialist judgments about when reflection was warranted: 52 of 60 interactions (86.7\%) were judged appropriate for follow-up, compared with 22 of 49 (44.9\%) in the technology probe. Noticing and Reasoning were rated highly in both deployments, suggesting that the main improvement was not in conducting reflection once initiated, but in better capturing specialist judgment about when reflection should occur.

\begin{table*}[t]
\caption{Post-study questionnaire results collected after Week~6 by condition. Values are mean ($SD$).}
\label{tab:poststudy-questionnaire}
\centering
\footnotesize
\renewcommand{\arraystretch}{1.15}
\setlength{\tabcolsep}{6pt}
\begin{tabularx}{\textwidth}{@{}Xcccc@{}}
\toprule
\textbf{Measure} &
\textbf{\dream} &
\textbf{Control} &
\textit{p} &
\textbf{Effect size} \\
\midrule
\textbf{Willingness to use} & \textbf{3.43 (1.19)} & \textbf{2.06 (0.84)} & \textbf{.015} & \textbf{$d=1.33$} \\
\textbf{Disclosure comfort} & \textbf{4.14 (0.93)} & \textbf{3.11 (1.14)} & \textbf{.044} & \textbf{$d=1.00$} \\
\textbf{Personal sleep insight} & \textbf{4.57 (0.61)} & \textbf{3.93 (0.35)} & \textbf{.027} & \textbf{$r_{rb}=.61$} \\
Sleep-data sensemaking & 4.29 (0.91) & 3.86 (0.38) & .059 & $r_{rb}=.47$ \\
\textbf{Motivation to improve sleep} & \textbf{4.21 (0.76)} & \textbf{3.36 (0.85)} & \textbf{.035} & \textbf{$d=1.06$} \\
\bottomrule
\end{tabularx}

\vspace{3pt}
\parbox{\textwidth}{\footnotesize
\textit{Note.} One-tailed Welch's $t$-tests were used for willingness, disclosure comfort, and motivation; one-tailed Wilcoxon rank-sum tests were used for personal sleep insight and sleep-data sensemaking. Effect sizes are Cohen's $d$ and rank-biserial correlation ($r_{rb}$). For reference, $|d|=.20/.50/.80$ and $|r_{rb}|=.10/.30/.50$ correspond approximately to small/medium/large effects. Bold indicates $p<.05$.

}
\end{table*}

\begin{figure*}[t]
    \centering
    \includegraphics[width=\textwidth]{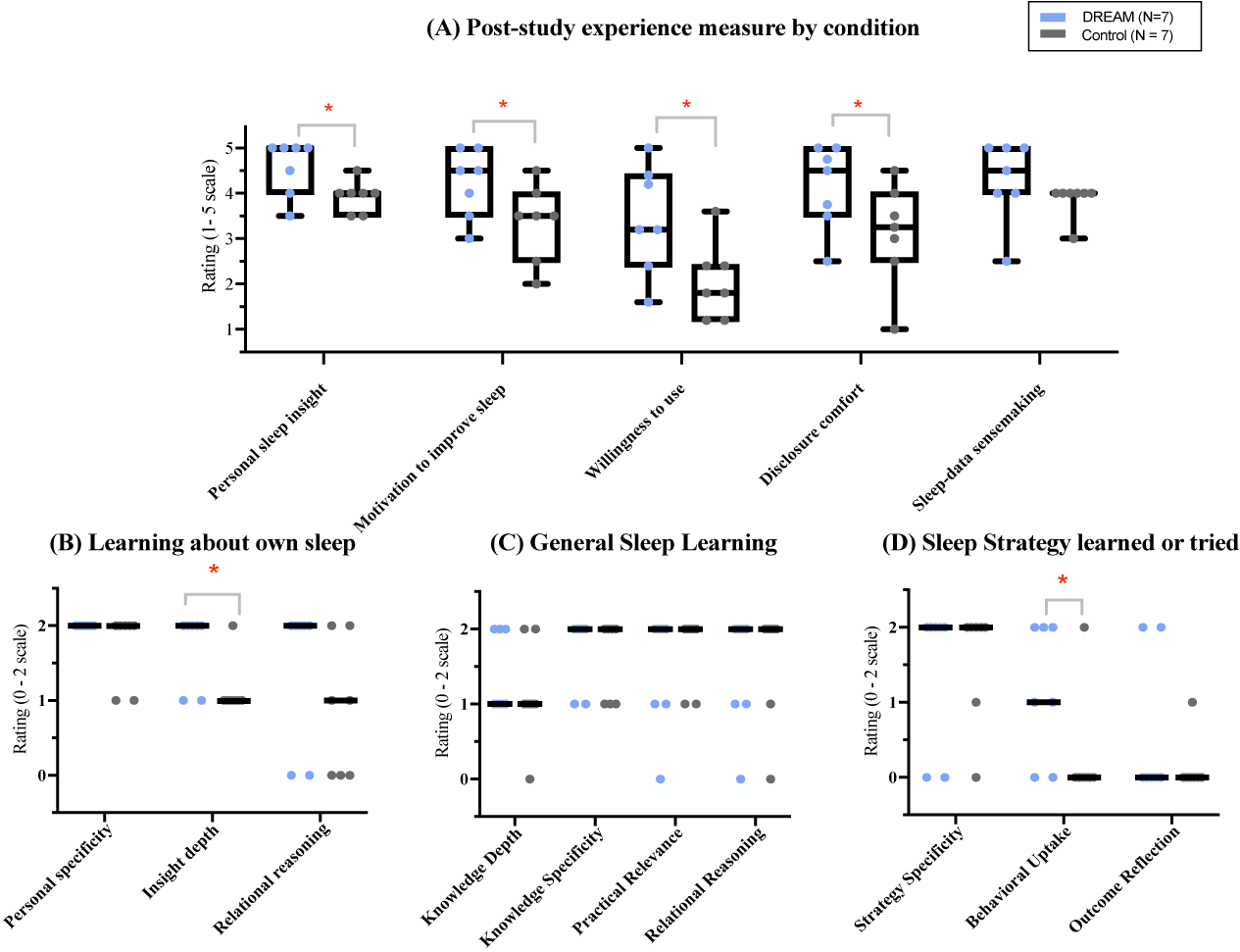}
    \caption{Post-study experience measures and coded open-ended responses by study condition. Points represent individual participants; boxes show the interquartile range with median center lines and min--max whiskers, while horizontal lines in the coded-response panels indicate medians. Experience measures used 1--5 scales and open-ended responses used 0--2 ordinal coding. Asterisks indicate between-condition differences with $p<.05$.}
    \Description{This figure contains four chart panels comparing two study conditions: a box plot across the top, labeled (A), and three dot plots side by side along the bottom, labeled (B), (C), and (D). Color identifies the condition throughout. Light blue dots represent participants in the DREAM condition, and dark grey dots represent participants in the control condition, as defined by a legend in the upper right reading ``DREAM (N=7)'' and ``Control (N = 7).'' For each measure, the DREAM data appear on the left and the control data on the right. A red asterisk above a light-grey bracket joining the two conditions marks a significant between-condition difference. Black lines show the summary statistics: box outlines, median lines, and whiskers in Panel A, and short thick horizontal bars in Panels B to D. Measure names are written diagonally beneath each x-axis. Panel A is titled in bold ``(A) Post-study experience measure by condition,'' and its y-axis is labeled ``Rating (1- 5 scale),'' running from 1 to 5. It shows five measures, each with a pair of thick black box plots overlaid with individual participant dots. For ``Personal sleep insight,'' the DREAM box spans approximately 4 to 5, with most dots at 5 and the lowest at 3.5. The control box spans approximately 3.5 to 4, with a median of 4 and dots between 3.5 and 4.5. For ``Motivation to improve sleep,'' the DREAM box spans approximately 3.5 to 5, with a median of about 4.5 and dots from 3 to 5. The control box spans approximately 2.5 to 4, with a median of about 3.5 and dots from 2 to 4.5. For ``Willingness to use,'' the DREAM box spans approximately 2.4 to 4.4, with a median of about 3.2 and dots from 1.6 to 5. The control box spans approximately 1.2 to 2.4, with a median of about 1.8 and dots from 1.2 to 3.6. For ``Disclosure comfort,'' the DREAM box spans approximately 3.5 to 5, with a median of about 4.5 and dots from 2.5 to 5. The control box spans approximately 2.5 to 4, with a median of about 3.3 and dots from 1 to 4.5. Each of these four measures carries a red asterisk. The fifth measure, ``Sleep-data sensemaking,'' has no asterisk. Its DREAM box spans approximately 4 to 5, with a median of about 4.5 and dots from 2.5 to 5. Six of the seven control dots sit at 4 and one sits at 3, so the control box collapses to a single line at 4. Panels B, C, and D share a y-axis labeled ``Rating (0 - 2 scale)'' with ticks at 0, 1, and 2. Individual dots sit at these three levels and are spread side by side when they share a level, and each condition has a black horizontal bar. Panel B is titled in bold ``(B) Learning about own sleep'' and shows three measures. For ``Personal specificity,'' all DREAM dots and the DREAM bar are at 2. Five control dots are at 2 and two are at 1, with the control bar at 2. ``Insight depth'' carries a red asterisk. Its DREAM bar is at 2, with most DREAM dots at 2 and two at 1. One control dot is at 2 and six are at 1, with the control bar at 1. For ``Relational reasoning,'' the DREAM bar is at 2, with most DREAM dots at 2 and two at 0. Two control dots are at 2, two at 1, and three at 0, with the control bar at 1. Panel C is titled in bold ``(C) General Sleep Learning'' and shows four measures, none marked with an asterisk. For ``Knowledge Depth,'' both bars are at 1, with a few dots from each condition at 2 and one control dot at 0. For ``Knowledge Specificity,'' both bars are at 2, with two DREAM dots and three control dots at 1. For ``Practical Relevance,'' both bars are at 2. Two DREAM dots are at 1 and one is at 0, and two control dots are at 1. For ``Relational Reasoning,'' both bars are at 2. Two DREAM dots are at 1 and one is at 0, and one control dot is at 1 and one is at 0. Panel D is titled in bold ``(D) Sleep Strategy learned or tried'' and shows three measures. For ``Strategy Specificity,'' both bars are at 2, with two DREAM dots at 0, one control dot at 1, and one control dot at 0. ``Behavioral Uptake'' carries a red asterisk. Three DREAM dots are at 2, the DREAM bar is at 1, and two DREAM dots are at 0. The control bar is at 0, with a single control dot at 2. For ``Outcome Reflection,'' both bars are at 0, with two DREAM dots at 2 and one control dot at 1. Together, the four panels illustrate that DREAM participants rated their experience higher than control participants on personal sleep insight, motivation to improve sleep, willingness to use, and disclosure comfort, while sleep-data sensemaking was similar across conditions. In the coded open-ended responses, DREAM participants showed greater insight depth about their own sleep and more behavioral uptake of sleep strategies, while general sleep learning was similar across the two conditions.}\label{fig:poststudy-findings}
\end{figure*}

\subsubsection{System Use and Experience}

The \dream condition received a mean SUS score of 69.29 ($SD=16.25$), slightly above the commonly used benchmark of 68 for digital health systems~\cite{hyzy2022system}, compared with 63.21 ($SD=21.25$) in the control condition. As shown in Table~\ref{tab:poststudy-questionnaire} and Fig.~\ref{fig:poststudy-findings}, \dream participants reported greater willingness to use the system ($p=.015$, $d=1.33$) and greater comfort disclosing information to it ($p=.044$, $d=1.00$). \dream interactions were also descriptively longer: participant-level median morning duration averaged 7.10 minutes ($SD=1.18$) for \dream participants and 4.99 minutes ($SD=0.90$) for control participants, as expected given the additional reflection turns in the \dream condition.

\begin{table*}[t]
\caption{Coded post-study open-ended responses collected after Week~6 by condition. Each criterion was coded from 0 to 2; values are mean ($SD$).}
\label{tab:openended-findings}
\footnotesize
\renewcommand{\arraystretch}{1.15}
\setlength{\tabcolsep}{5pt}
\centering
\begin{tabularx}{\textwidth}{@{}lXcccc@{}}
\toprule
\textbf{Question} & \textbf{Criterion} & \textbf{\dream} & \textbf{Control} & \textit{p} & \textbf{$r_{rb}$} \\
\midrule
\multirow{3}{*}{Learning about own sleep}
& Personal specificity & 2.00 (0.00) & 1.71 (0.49) & .086 & .29 \\
& \textbf{Insight depth} & \textbf{1.71 (0.49)} & \textbf{1.14 (0.38)} & \textbf{.022} & \textbf{.57} \\
& Relational reasoning & 1.43 (0.98) & 0.86 (0.90) & .131 & .35 \\
\midrule
\multirow{4}{*}{General sleep learning}
& Knowledge depth & 1.43 (0.53) & 1.14 (0.69) & .233 & .22 \\
& Specificity & 1.71 (0.49) & 1.57 (0.53) & .323 & .14 \\
& Practical relevance & 1.43 (0.79) & 1.71 (0.49) & .775 & $-.18$ \\
& Relational reasoning & 1.43 (0.79) & 1.57 (0.79) & .700 & $-.12$ \\
\midrule
\multirow{3}{*}{Sleep strategy learned or tried}
& Strategy specificity & 1.43 (0.98) & 1.57 (0.79) & .596 & $-.04$ \\
& \textbf{Behavioral uptake} & \textbf{1.14 (0.90)} & \textbf{0.29 (0.76)} & \textbf{.036} & \textbf{.53} \\
& Outcome reflection & 0.57 (0.98) & 0.14 (0.38) & .238 & .18 \\
\bottomrule
\end{tabularx}

\vspace{3pt}
\parbox{\textwidth}{\footnotesize
\textit{Note.} Between-condition comparisons used one-tailed Wilcoxon rank-sum tests. Positive $r_{rb}$ values indicate higher scores in the \dream condition. Question wording and coding criteria are given in Table~\ref{tab:openended-coding}.
}
\end{table*}

\subsubsection{Sleep Understanding and Action}

\paragraph{Personal Sleep Understanding.}
\dream participants reported greater personal sleep insight ($p=.027$, $r_{rb}=.61$; Table~\ref{tab:poststudy-questionnaire}). Coded responses showed a similar pattern: \dream participants received higher insight-depth scores when describing what they had learned about their own sleep ($p=.022$, $r_{rb}=.57$; Table~\ref{tab:openended-findings}), reflecting greater interpretation of sleep patterns in relation to their behaviors or circumstances. Personal specificity and relational reasoning did not differ significantly between conditions.

\paragraph{Sleep Knowledge.}
Coded responses about general sleep learning showed no significant between-condition differences across knowledge depth, specificity, practical relevance, or relational reasoning. Repeated measures likewise showed no condition-specific differences in Sleep Beliefs or SHAPS knowledge. Across both conditions, however, sleep-hygiene awareness increased over time ($F(2,23.5)=3.76$, $p=.038$), as did caffeine knowledge ($F(2,24.1)=10.53$, $p<.001$), indicating improvement across participants regardless of condition. Directional pre-to-post contrasts showed no significant between-condition differences. Full longitudinal results are reported in Appendix~\ref{app:distal-outcomes}.

\paragraph{Motivation and Uptake.}
\dream participants reported greater motivation to improve their sleep ($p=.035$, $d=1.06$; Table~\ref{tab:poststudy-questionnaire}). Their open-ended responses also received higher behavioral-uptake scores ($p=.036$, $r_{rb}=.53$; Table~\ref{tab:openended-findings}), indicating stronger evidence of considering or trying sleep-related strategies. Strategy specificity and outcome reflection did not differ significantly between conditions. Sleep Self-Efficacy likewise showed no clear condition-specific change over the deployment, and the directional pre-to-post contrast was not significant.

\subsubsection{Sleep and Behavioral Outcomes}
Sleep and behavioral outcomes were similar across conditions. We observed no significant condition-specific differences in PSQI, SHAPS sleep-hygiene practices, nightly sleep measures, or routinely recorded sleep-related behaviors over the six-week deployment. Across both conditions, total sleep time (TST) and sleep efficiency (SE) increased over time, while wake after sleep onset (WASO) decreased. Other nightly sleep measures and routinely recorded behaviors, including late caffeine use and pre-bed screen use, showed no clear changes over time. Full longitudinal results are reported in Appendix~\ref{app:distal-outcomes}.

\subsubsection{Participant Perspectives}
Thematic analysis identified five related aspects of participants' experiences that contextualize the findings above:

\paragraph{Attending to Sleep.}
Participants in both conditions described attending more closely to their sleep through repeated diary use. \texttt{CTRL-P12}, for example, inferred from questions about medication, caffeine, food, and exercise that \myc{\textit{``if I'm being asked all these things, they are related to sleep,''}} which \myc{\textit{``really made me more aware.'}} Similarly, \texttt{\dreamNospace-P14} observed, \myc{\textit{``In the past, I would pay attention to how long I slept\dots\ I didn't really pay attention to when I go to bed and when I wake up before.''}} These accounts suggest that repeated monitoring itself directed attention toward sleep behaviors and measures participants had previously overlooked, consistent with the increases in sleep-hygiene awareness observed across both conditions.

\paragraph{Interpreting Changes.}
The conditions differed more clearly in what happened after participants noticed aspects of their sleep. Control participants wanted greater support in identifying meaningful changes and understanding what those changes might mean. \texttt{CTRL-P10} wanted \myc{\textit{``explanations as to like if things I was doing was negatively impacting my sleep that day,''}} while \texttt{CTRL-P11} imagined the system drawing attention to a pattern by saying, \myc{\textit{``notice this thing. Like here's a comment.''}} \texttt{CTRL-P01} similarly wanted the system to aggregate entries, for example, \myc{\textit{``I've noticed you've been going to sleep late.''}}

\dream participants instead described developing explanations themselves. When \texttt{\dreamNospace-P13} was asked why he had stayed in bed longer than usual, he connected the change to recently starting pickleball and going to the gym. \texttt{\dreamNospace-P08} similarly recalled being asked "why do you think that happened?" after a change in his time in bed, which led him to identify phone use as a possible explanation. \texttt{\dreamNospace-P07} valued that these conversations could \myc{\textit{``go back to the previous diary,''}} allowing a current observation to be considered alongside earlier context. These accounts align with the greater personal sleep insight reported by \dream participants (Fig.~\ref{fig:poststudy-findings}). Thus, \textbf{while monitoring supported attention in both conditions, guided reflection provided an additional step for connecting an observed change to participants' own circumstances}. 

\paragraph{Timing Reflection Selectively.}
Reflection was valued when the selected change felt meaningful and questioned when it did not. \texttt{\dreamNospace-P16} valued that on a regular day \myc{\textit{``it would not ask much,''}} whereas \texttt{\dreamNospace-P14} noted that \myc{\textit{``when there's no issue, it'll be nice to not try to find a difference.''}} \texttt{\dreamNospace-P07} similarly noted that a difference of half an hour could simply reflect waking earlier for work. These accounts reinforce \textbf{the importance of distinguishing detectable change from a useful opportunity for reflection.}

\paragraph{Making Education Personally Relevant.}
Participants in both conditions sometimes found sleep education useful, but differed in how directly that information connected to their own sleep. Generic education (Control) was useful when it happened to align with participants' needs. \texttt{CTRL-P05}, for example, found a suggestion about not going to bed hungry useful but described other tips as \myc{\textit{``something I already knew''}} and \myc{\textit{``not very specific to my sleep.''}} \dream participants instead described education that followed issues they had already considered in relation to their own sleep. \texttt{\dreamNospace-P14} reconsidered how long caffeine could affect her sleep, while \texttt{\dreamNospace-P16} discussed her phone use before receiving screen-time guidance.

\paragraph{Developing a Rationale for Action.}
Participants also differed in how they described moving from receiving sleep education to trying a strategy. Control participants acted on a generic tip when it independently appeared applicable. \dream participants more often described actions as following from an interpretation they had first developed. \texttt{\dreamNospace-P14} reduced caffeine later in the day, \texttt{\dreamNospace-P16} regulated screen time and reported that \myc{\textit{``it did help,''}} and \texttt{\dreamNospace-P13} described moving toward knowing \myc{\textit{``how to take action and what strategies to implement.''}} These accounts are consistent with the higher motivation and behavioral uptake reported by \dream participants.

\section{Discussion}

We synthesize findings from the technology probe, the specialist reviews, and the field study to consider what guided reflection contributed to everyday sleep tracking, how the resulting understanding related to action, and what our design process suggests for LLM-supported reflection in personal health more broadly.

\subsection{Making Sleep Data Personally Meaningful through Guided Reflection}
Repeated tracking and education supported general awareness in both conditions. Participants in both groups described attending to bedtimes, wake times, and daily behaviors they had previously overlooked, and sleep-hygiene awareness and caffeine knowledge improved over the deployment regardless of condition. What tracking and generic education did not provide was help interpreting why a particular night looked different. Control participants wanted the system to notice patterns, point out meaningful changes, and help them understand whether something they had recorded might have contributed to that night's sleep. General knowledge about factors that influence sleep could tell participants what matters in principle, but not why a particular change may have occurred for them. This reflects a longstanding challenge in self-tracking: collecting data can support awareness, but does not necessarily help people turn those observations into personally meaningful interpretations~\cite{choe2014quantified,ravichandran2017making,purnell2023wearable}.

Guided reflection addressed this gap by changing how personalized guidance was reached. Control participants received a sleep-hygiene tip after every morning diary, whereas \dream participants received education only when reflection established an actionable contributor. Despite receiving education less frequently, \dream participants reported more personal sleep insight, described what they had learned about their own sleep in greater depth, and were more willing to continue using the system. Importantly, \dream did not move directly from an observed change to an explanation or recommendation. Instead, it brought the observation into conversation and first asked what the participant thought had contributed~\cite{padesky1993socratic,overholser2011guided}, introducing recorded evening context only as a cue when the participant remained uncertain. Participants described making these connections themselves, for example by linking longer time in bed to a new exercise routine (\texttt{\dreamNospace-P13}), identifying phone use as a possible contributor (\texttt{\dreamNospace-P08}), or relating a current observation to an earlier diary entry (\texttt{\dreamNospace-P07}). These findings suggest that guided reflection supported participants in doing the interpretive work themselves, rather than having the system infer the meaning of their data on their behalf. \textbf{By not simply off-loading this cognitive effort to the system, participants became active participants in identifying the causes of changes in their sleep}. This active engagement helps them learn more about sleep science and recognize what they can actually act on, guiding them to begin the path toward sleep behavior change.


\dream participants also reported higher comfort disclosing information to the system. Because both conditions used the same voice-based diaries, this pattern is unlikely to reflect interaction modality alone. One possibility is that reflection changed how participants perceived the value of what they disclosed. In the control condition, information recorded in the diary did not visibly shape the education that followed, whereas in the \dream condition, participants' responses influenced what the system asked next and whether education was provided. Disclosure may therefore feel more meaningful when users can see that the context they provide shapes the interaction.

\subsection{Connecting Personal Understanding to Action}
In the control condition, generic tips arrived without any connection to the participant's own sleep, so participants had to work out for themselves whether a tip applied to them, and they often described tips as already known or \myc{\textit{"not very specific to my sleep."}} In the \dream condition, education followed only after participants had connected a sleep change to something in their own lives, so the reason for a strategy was already established before it was offered. This reflects guided discovery, where interpretations developed by the individual can be more meaningful and actionable than those provided to them~\cite{padesky1993socratic,overholser2011guided}. \dream participants' higher motivation and reported uptake, including \texttt{\dreamNospace-P13}'s description of moving from recognizing patterns to knowing \myc{\textit{"how to take action and what strategies to implement,"}} are consistent with this. Personalization therefore involved not only tailoring what guidance was given, but grounding that guidance in an actionable contributor the participant had already identified, similar to how behavioral strategies in CBT-I are developed from a patient's account of their sleep~\cite{walker2022cbti,edinger2021guideline}. 
Our findings suggest that \textbf{guided reflection may support an earlier step toward behavior change by helping users understand why a strategy is relevant to their own sleep}~\cite{li2010stage,epstein2015lived}.

\subsection{Design Implications for LLM-Supported Reflection in Personal Health}
Our findings suggest that the value of an LLM in personal health coaching lies not only in generating tailored content, but in facilitating how users interpret their own data. Earlier reflection systems relied on predefined prompts or visualizations that left interpretation to the user~\cite{liang2016sleepexplorer,liu2020slumberbot,kocielnik2018reflection}, while recent LLM-based systems have largely used models to produce narratives, feedback, or recommendations~\cite{stromel2024narrating,loerakker2025giveandtake,wang2025healthguru}. \dream instead presented an observation, invited the user to consider what might have contributed, followed their account with clarifying questions, and only then determined what guidance, if any, should follow. 

This interaction pattern could extend beyond sleep. Physical activity, diet, and stress tracking similarly produce records whose meaning depends on circumstances the data do not capture~\cite{li2010stage,epstein2015lived,choe2014quantified}. An LLM-powered coach could ask what surrounded a period of reduced activity or a change in eating patterns before suggesting a strategy, so that guidance follows from a contributor the user has identified rather than from the data alone. Because guided discovery originates in cognitive behavioral therapy, the same principle may also extend to behavioral health contexts such as stress or anxiety, provided that domain experts define what observations warrant reflection, what questions are appropriate to ask, and what guidance can safely follow. More broadly, we argue that \textbf{LLMs should move beyond traditional content personalization to tailor the process through which users make sense of their own data and decide what to do next.}

\subsection{Translating Specialist Judgment into Adaptive System Logic}
Our design process shows that translating specialist judgment in clinical practice into an LLM-supported system involves more than transferring clinical knowledge or conversational techniques. It requires identifying the different forms of judgment that make the practice work. Guided discovery depends on following a person's account, clarifying what they mean, and adapting the next question to what they say, which LLMs can support more flexibly than predefined questions or scripts~\cite{liu2020slumberbot,kocielnik2018reflection,fitzpatrick2017woebot}. But specialist practice also involves judgments about whether an observation is worth raising at all, how much evidence is enough to pursue it, and when no explanation or guidance is needed. These judgments depended on baseline, recurrence, and relationships across measures, and were not captured by documented clinical thresholds or sleep knowledge alone. Translating the practice therefore meant making these otherwise implicit distinctions explicit in the system, while also preserving a central restraint of guided discovery.


The iterative specialist review was important to this translation because many of these distinctions emerged only after the system was used. Simulated cases helped establish an initial representation of the practice, but the technology probe exposed judgments that had not yet been represented~\cite{hutchinson2003technologyprobes}. Specialists judged only 44.9\% of flagged interactions as appropriate for follow-up. After incorporating these judgments into the refined system, 86.7\% of reviewed interactions were judged appropriate for follow-up. We therefore see \textbf{the involvement of domain experts not as a one-time transfer of expertise into prompts or rules, but as a process of progressively making tacit judgment explicit and testable}. More broadly, translating expert practices into LLM-supported health systems may require designing for both what the model can flexibly do and the judgments that must structure when, why, and how that capability is used.

\subsection{Limitations and Future Work}
Our study has several limitations. First, our sample was small and primarily drawn from a university population, limiting the demographic diversity of participants and the generalizability of our findings. Future deployments should evaluate the system with larger and more diverse populations to examine how guided reflection operates across different daily routines and tracking needs.

Second, implementing the system as a custom Amazon Alexa skill introduced usability constraints, including speech-recognition errors and cases where Alexa interpreted participants' speech as another command, requiring them to repeat or restart the interaction. These issues could disrupt longer reflection conversations and may have affected participants' engagement. Future work could explore guided reflection on more flexible conversational platforms, including mobile interfaces that support both voice and text.

Third, the field study was not designed to isolate the effect of guided reflection from other aspects of personalization. Participants in the Generic Education condition received education independent of their sleep data, whereas \dream combined guided reflection with education tailored to the issue surfaced through the interaction. Differences between conditions therefore cannot be attributed specifically to the reflective process. Future studies could disentangle these components by comparing guided reflection with personalized education delivered without reflection, or by experimentally varying reflection and education independently.

Finally, although we constrained educational content to specialist-provided sleep-hygiene material and iteratively refined the system with sleep specialists, \dream did not include explicit mechanisms for detecting hallucinated, unsupported, or clinically inappropriate responses. Future LLM-supported health systems should incorporate stronger safeguards, enforce clear scope boundaries, and provide pathways to appropriate clinical support when concerns exceed the system's intended role.

\section{Conclusion}
We presented \dream, an LLM-powered, voice-based system that adapts guided discovery from behavioral sleep medicine to everyday sleep tracking. The system selectively brings changes in a user's sleep record into conversation, supports the user's interpretation of those changes, and grounds subsequent education in that understanding. Through a six-week field study with 14 participants, we found that \dream supported deeper personal understanding of sleep and a clearer rationale for the strategies participants tried, compared to a generic-education control. Both conditions improved participants' general sleep knowledge, but that knowledge did not by itself tell participants why their own sleep had changed. Guided reflection provided that interpretive step, giving participants a personal rationale for acting on subsequent guidance. Our findings highlight an opportunity to design behavior-change technologies that use personal data not only to personalize guidance, but also to support the personal understanding that makes that guidance meaningful and actionable.

\section*{Declaration of generative AI and AI-assisted technologies in the writing process}

During the preparation of this work, the authors used ChatGPT and Claude in order to cut down repetitions and improve readability and language. Figures were designed with the help of Claude and figure descriptions were generated using Claude. After using these tools/services, the authors reviewed and edited the content as needed and take full responsibility for the content of the publication. 

\newpage

\bibliographystyle{ACM-Reference-Format}
\bibliography{references}
\newpage
\appendix
 
\section{System Implementation Details}
\label{app:implementation}
 
\subsection{Platform, Architecture, and State Management}
We implemented DREAM as an LLM-powered voice application on the Amazon Alexa platform. An AWS Lambda function handled incoming Alexa requests and communication with the middleware, while the backend logic was hosted on Microsoft Azure. To accommodate LLM processing within Alexa's response-time constraints, we adopted the dual-API middleware architecture introduced by Mahmood et al.~\cite{mahmood2025user}, which coordinated Alexa interactions with backend processing and provided brief conversational responses while longer-running operations were completed. OpenAI's GPT-5.2 supported the system's conversational and adaptive components. Structured diary records, derived sleep metrics, and interaction histories were stored separately for each participant in participant-specific Google Sheets, which served as the system's database for longitudinal tracking.

\subsection{Prompting Strategy and Module Handlers}
Because the system required different conversational behaviors at different stages, we implemented distinct LLM handlers, each driven by its own specific system prompt:
 
\subsubsection{Diary Intake Conversation Handler}
A dedicated prompt instructed the LLM to conduct morning and evening diaries by asking predefined questions (e.g., bedtime, sleep onset latency, mood, caffeine) one at a time in conversational language, allowing clarification when responses were ambiguous.
 
\subsubsection{Text-to-JSON Converter}
After each diary conversation, a separate prompt instructed the LLM to function as a data extractor, converting the natural language transcript into predefined structured JSON fields for backend metric derivation.
 
\subsubsection{Reflection Handler}
When the backend flagged a change, a distinct reflection prompt was triggered, injecting the specific sleep deviation and relevant evening context into the LLM context window. The prompt strictly constrained the LLM to follow the specialist-informed decision tree (asking open-ended questions first and offering evening context only as cues rather than asserted causes).
 
\subsubsection{Education Handler}
This prompt provided the LLM with specialist-approved sleep-hygiene guidelines, instructing the model to synthesize a brief response connecting the user's articulated interpretation to one or two actionable strategies, strictly forbidding diagnostic claims outside the provided text.

\section{Sleep Diary Questions}
\label{app:sleep-diary-questions}
 
Participants completed conversational morning and evening sleep diaries. The
required diary content was fixed, while the LLM phrased each question
conversationally and could ask clarification questions when responses were
ambiguous or incomplete.
 
\begin{table}[h!]
\caption{Information collected in the morning and evening conversational sleep diaries.}
\label{tab:sleep-diary-questions}
\footnotesize
\renewcommand{\arraystretch}{1.15}
\setlength{\tabcolsep}{5pt}
\centering
\begin{tabularx}{\textwidth}{@{}p{2.5cm}X@{}}
\toprule
\textbf{Diary} & \textbf{Information Collected} \\
\midrule
 
\textbf{Morning}
&
Time of getting into bed; lights-out time; sleep onset latency (SOL);
planned-bedtime adherence; number of nighttime awakenings (NWAK);
total time awake after sleep onset (WASO); final awakening time;
out-of-bed time; deviation from planned wake time; subjective sleep quality;
sleep-aid use and timing; and other disruptions or circumstances affecting
the previous night's sleep. \\
 
\midrule
 
\textbf{Evening}
&
Mood; stress; fatigue; pain; naps and their timing and duration;
caffeine and alcohol intake, including amount and timing; exercise and
physical activity; medication use; evening and pre-bed activities;
sleep-environment factors such as light, noise, and temperature; and other
events or circumstances that might be relevant to sleep. \\
 
\bottomrule
\end{tabularx}
\end{table}

\section{Open-Ended Response Coding Criteria}
\label{app:openended-coding}
 
Responses were de-identified and presented without condition labels before coding. Two researchers independently coded three responses per question, achieving agreement on 29 of 30 coding judgments. Disagreements were resolved through discussion, after which the remaining responses were divided equally between the two coders.
 
\begin{table}[h!]
\caption{Coding criteria for the three post-study open-ended questions. Each criterion was coded on a 0--2 ordinal scale, with higher scores indicating stronger evidence of the corresponding quality.}
\label{tab:openended-coding}
\footnotesize
\renewcommand{\arraystretch}{1.15}
\setlength{\tabcolsep}{4pt}
\centering
\begin{tabularx}{\textwidth}{@{}p{3.2cm}p{3.0cm}X@{}}
\toprule
\textbf{Question} & \textbf{Criterion} & \textbf{Coding Interpretation} \\
\midrule
\textbf{What did you learn about your own sleep?}
& Personal specificity
& 0 = no reference to the participant's own sleep; 1 = a general reference to their sleep without specific measures or circumstances; 2 = specific aspects of their own sleep, such as particular measures, times, or situations. \\
& Insight depth
& 0 = restates a fact or observation without interpretation; 1 = notes a pattern in their sleep without explaining what may underlie it; 2 = interprets a pattern in relation to their own behaviors or circumstances. \\
& Relational reasoning
& 0 = no relationship described; 1 = asserts a relationship between a behavior and sleep without elaboration; 2 = describes how a behavior or circumstance related to a specific aspect of their sleep. \\
\midrule
\textbf{What did you learn about sleep more generally?}
& Knowledge depth
& 0 = no sleep-related knowledge reported; 1 = states a general fact about sleep; 2 = states a fact together with a mechanism or condition under which it applies. \\
& Specificity
& 0 = a vague or generic statement about sleep; 1 = identifies a specific factor affecting sleep; 2 = identifies a specific factor with detail such as timing, amount, or affected measure. \\
& Practical relevance
& 0 = no actionable implication; 1 = implies an action without specifying it; 2 = states an actionable behavior or strategy following from the knowledge. \\
& Relational reasoning
& 0 = no relationship described; 1 = asserts a relationship between a factor and sleep without elaboration; 2 = describes how a factor affects a specific aspect of sleep. \\
\midrule
\textbf{Which sleep-related strategies did you learn or try?}
& Strategy specificity
& 0 = no strategy named; 1 = names a strategy in general terms; 2 = names a strategy with specific parameters such as timing, amount, or context. \\
& Behavioral uptake
& 0 = no attempt described; 1 = considered or intended to try the strategy; 2 = reports having tried the strategy. \\
& Outcome reflection
& 0 = no outcome mentioned; 1 = states an outcome without detail; 2 = describes the outcome and relates it to a specific aspect of their sleep. \\
\bottomrule
\end{tabularx}
\end{table}

\section{Post-Study Questionnaire Items}
\label{app:poststudy-items}
 
Table~\ref{tab:poststudy-items} reports the custom post-study questionnaire items used to construct the participant-reported measures analyzed in the field study.
 
\begin{table}[h]
\caption{Custom post-study questionnaire measures and items. All items were rated on 5-point Likert scales.}
\label{tab:poststudy-items}
\footnotesize
\renewcommand{\arraystretch}{1.15}
\setlength{\tabcolsep}{5pt}
\centering
\begin{tabularx}{\textwidth}{@{}p{3.0cm}Xc@{}}
\toprule
\textbf{Measure} & \textbf{Items} & \textbf{$\alpha$} \\
\midrule
 
Willingness to use
&
I would want to continue using this system.\newline
I would recommend this system to someone who wants to track/improve their sleep.\newline
I could see myself using a system like this regularly in daily life.\newline
I intend to use this assistant in the future.\newline
If available, I would choose to continue using this assistant.
& .93 \\
\midrule
 
Disclosure comfort
&
I felt comfortable sharing personal details during the diary/coaching interactions.\newline
It was easy to share details about my sleep habits and routines with the system.\newline
I felt like I could be personal when talking with the system.\newline
I felt comfortable with how much personal information I shared through the diary and coaching conversations.
& .87 \\
\midrule
 
Personal sleep insight
&
The system gives me ideas about what I could try to improve my sleep.\newline
I have a good understanding of what affects my own sleep.
& .82 \\
\midrule
 
Sleep-data sensemaking
&
It was easy to make sense of changes in my sleep using the system.\newline
The system made it easy to get an overview of my sleep over time.
& .71 \\
\midrule
 
Motivation to improve sleep
&
This experience increased my motivation to make changes to improve my sleep.\newline
I feel motivated to try new strategies (e.g., a consistent bedtime, limiting screens) to sleep better.
& .83 \\

\bottomrule
\end{tabularx}
\end{table}

\section{Sleep and Behavioral Outcomes}
\label{app:distal-outcomes}
 
Full longitudinal results for questionnaire and nightly sleep outcomes are reported below.
 
\begin{table}[h!]
\caption{Longitudinal results for repeated questionnaire measures. Condition $\times$ Time tests indicate whether trajectories differed between conditions; Time tests indicate change across participants regardless of condition.}
\label{tab:repeated-questionnaires}
\footnotesize
\renewcommand{\arraystretch}{1.15}
\setlength{\tabcolsep}{6pt}
\centering
\begin{tabularx}{\textwidth}{@{}p{4.2cm}ccl X@{}}
\toprule
\textbf{Measure} &
\textbf{Condition $\times$ Time} &
\textit{p} &
\textbf{Time} &
\textbf{Summary} \\
\midrule
 
PSQI &
$F(2,24.6)=1.79$ & .189 &
$p>.05$ &
No significant condition-specific or overall time effect. \\
 
Sleep Self-Efficacy &
$F(2,23.7)=3.40$ & .051 &
$p>.05$ &
No significant condition-specific or overall time effect. \\
 
Sleep Beliefs &
$F(2,21.5)=0.76$ & .480 &
$p>.05$ &
No significant condition-specific or overall time effect. \\
 
SHAPS: sleep-hygiene awareness &
$F(2,23.4)=0.39$ & .684 &
$F(2,23.5)=3.76$, $p=.038$ &
Increased over time across conditions. \\
 
SHAPS: caffeine knowledge &
$F(2,23.9)=0.20$ & .818 &
$F(2,24.1)=10.53$, $p<.001$ &
Increased over time across conditions. \\
 
SHAPS: sleep-hygiene practices &
$F(2,23.8)=1.31$ & .289 &
$p>.05$ &
No significant condition-specific or overall time effect. \\
 
\bottomrule
\end{tabularx}
 
\vspace{3pt}
\parbox{\textwidth}{\footnotesize
\textit{Note.} No questionnaire showed a significant condition-by-time
interaction. Directional pre-to-post contrasts likewise showed no significant
between-condition differences in change (all $p>.05$).
}
\end{table}
 
\begin{table}[h!]
\caption{Longitudinal model results for nightly sleep outcomes.}
\label{tab:nightly-sleep-outcomes}
\footnotesize
\renewcommand{\arraystretch}{1.15}
\setlength{\tabcolsep}{6pt}
\centering
\begin{tabularx}{\textwidth}{@{}Xccc@{}}
\toprule
\textbf{Measure} &
\textbf{Condition $\times$ Week ($p$)} &
\textbf{Main Effect of Week ($p$)} &
\textbf{Direction} \\
\midrule
TST & .818 & .005 & Increased \\
SOL & .381 & $>.05$ & --- \\
WASO & .741 & .006 & Decreased \\
SE & .561 & .040 & Increased \\
Sleep quality & .890 & $>.05$ & --- \\
TIB & .736 & $>.05$ & --- \\
\bottomrule
\end{tabularx}
 
\vspace{3pt}
\parbox{\textwidth}{\footnotesize
\textit{Note.} Direction is shown for significant main effects of study week.
}
\end{table}
 
\begin{table}[H]
\caption{Exploratory longitudinal changes in routinely recorded sleep-related behaviors.}
\label{tab:behavioral-outcomes}
\footnotesize
\renewcommand{\arraystretch}{1.15}
\setlength{\tabcolsep}{6pt}
\centering
\begin{tabularx}{\columnwidth}{@{}Xc@{}}
\toprule
\textbf{Behavior} & \textbf{Between-Condition Comparison ($p$)} \\
\midrule
Late caffeine use & .928 \\
Alcohol use & .498 \\
Long naps ($>30$ min) & .604 \\
Exercise & .488 \\
Pre-bed screen use & .259 \\
Planned-bedtime adherence & .391 \\
\bottomrule
\end{tabularx}
 
\vspace{3pt}
\parbox{\columnwidth}{\footnotesize
\textit{Note.} For each behavior, participant-level longitudinal change was compared between conditions using an exact randomization test. No between-condition difference reached statistical significance.
}
\end{table}

\section{Detailed Specialist Evaluation}
\label{app:specialist-evaluation}
 
\begin{table}[h!]
\caption{Specialist evaluation of system interactions in the technology probe and field study. Stage-level ratings used 5-point scales, with higher scores indicating more favorable assessments. Values are mean (\textit{SD}).}
\label{tab:specialist-ratings}
\centering
\footnotesize
\renewcommand{\arraystretch}{1.15}
\setlength{\tabcolsep}{4pt}

\begin{tabular*}{\textwidth}{@{\extracolsep{\fill}}llcccc@{}}
\toprule
& & \multicolumn{2}{c}{\textbf{Technology Probe}}
& \multicolumn{2}{c}{\textbf{Field Study}} \\
\cmidrule(lr){3-4}\cmidrule(lr){5-6}
\textbf{Stage} & \textbf{Criterion}
& \textbf{\textit{n}} & \textbf{Result / \textit{M} (\textit{SD})}
& \textbf{\textit{n}} & \textbf{Result / \textit{M} (\textit{SD})} \\
\midrule
 
Flagging
& Appropriate for reflection
& 49 & 22 (44.9\%)
& 60 & 52 (86.7\%) \\
\midrule
 
\multirow{4}{*}{Noticing}
& Accuracy        & 26 & 4.92 (0.27) & 52 & 4.88 (0.38) \\
& Appropriateness & 26 & 4.85 (0.61) & 52 & 4.88 (0.43) \\
& Relevance       & 26 & 4.81 (0.57) & 52 & 4.88 (0.38) \\
& Specificity     & 26 & 4.88 (0.43) & 52 & 4.98 (0.14) \\
\midrule
 
\multirow{4}{*}{Reasoning}
& Accuracy        & 24 & 4.50 (0.78) & 52 & 4.37 (1.05) \\
& Appropriateness & 24 & 4.58 (0.83) & 52 & 4.46 (0.94) \\
& Relevance       & 24 & 4.54 (0.93) & 52 & 4.35 (1.05) \\
& Specificity     & 24 & 4.83 (0.38) & 52 & 4.44 (0.92) \\
\midrule
 
\multirow{4}{*}{Action}
& Accuracy        & 25 & 3.64 (1.22) & 41 & 4.24 (0.70) \\
& Appropriateness & 25 & 3.96 (1.17) & 41 & 3.78 (0.94) \\
& Relevance       & 25 & 3.84 (1.18) & 41 & 3.63 (0.99) \\
& Specificity     & 24 & 4.67 (0.56) & 41 & 4.27 (0.78) \\
\bottomrule
\end{tabular*}
 
\vspace{3pt}
\parbox{\textwidth}{\footnotesize
\textit{Note.} Flagging was evaluated across all reviewed interactions. For stage-level evaluations, \textit{n} indicates the number of available specialist ratings contributing to each estimate and therefore varies where ratings were unavailable. In the field study, Noticing and Reasoning were evaluated for interactions judged appropriate for follow-up, while Action was evaluated only for interactions that proceeded to education.
}
\end{table}
 

\end{document}